\documentclass[useAMS,usenatbib,usegraphicx]{mn2e}
\def \zemax{{\it\bf Zemax}}
\def \zo{zeroth order\ }
\def \uv{UV\ }
\def \uvg{{\uv grism}\ }
\def \ucg{{\uv grism clocked mode}\ }
\def \ung{{\uv grism nominal mode}\ }
\def \ucG{{\uv grism clocked mode}}
\def \unG{{\uv grism nominal mode}}
\def \visc{{visible grism clocked mode}\ }
\def \visn{{visible grism nominal mode}\ }
\def \visC{{visible grism clocked mode}}
\def \visN{{visible grism nominal mode}}
\def \visg{{visible grism}\ }
\def \uvG{{\uv grism}}
\def \visG{{visible grism}}
\def \ungb{{\uv nominal}\ }
\def \ucgb{{\uv clocked}\ }
\def \viscb{{visible clocked}\ }
\def \visnb{{visible nominal}\ }
\def \Sw{{\it Swift}\ }
\def \xmm{{\it XMM}\ }
\def \OM{{\it XMM}-OM}
\def \iue{{\it IUE}}
\def \uvot{UVOT\ }
\def \uvotpy{{\tt UVOTPY} software}

\def \ftool{{\em ftool}\ }
\def \CALDB {{\it Swift} {\em CALDB}}
\def \lam{{$\lambda$}}
\def \fits{{FITS\ }}
\def \mod8{MOD-8}
\def \changed{}

\title[The Swift-UVOT ultraviolet and visible grism calibration]
   {\it\bf Calibration of the Swift-UVOT ultraviolet and visible grisms} 

\author[The Swift UVOT Team]
  {N.~P.~M. Kuin$^1$\thanks{email: n.kuin@ucl.ac.uk}, 
     W.~Landsman$^2$, 
     A.~A.~Breeveld$^1$, 
     M.~J.~Page$^1$,  
     H.~Lamoureux$^1$,
          \newauthor  
     C.~James$^1$, 
     M.~Mehdipour$^1$,
     M.~Still$^3$,   
     V.~Yershov$^1$,
     P.~J.~Brown$^5$, 
     M.~Carter$^1$,
     \newauthor  
      K. O. Mason$^6$, 
     T.~Kennedy$^1$,
      F.~Marshall$^7$,
      P.~W.~A.~Roming$^{4,8,10}$,
      M.~Siegel$^4$,
     \newauthor  
      S.Oates$^{1,11}$, 
      P.~J.~Smith$^1$,  
      and M.~De~Pasquale$^{1,9}$
      \\   
      \\
$^1$Mullard Space Science Laboratory/UCL, 
   Holmbury St. Mary, Dorking, Surrey, RH5 6NT, UK\\
$^2$Space Telescope Science Institure, Baltimore, MD 00000, USA\\ 
$^3$NASA Ames Research Center, M/ Table S 244-40, Moffett Field, CA 94035, USA \\
$^4$Department of Astronomy \& Astrophysics, Penn State University, 
    525 Davey Laboratory, University Park, PA 16802, USA\\
$^5$George P. and Cynthia Woods Mitchell Institute for Fundamental Physics \& Astronomy,
Texas A. \& M. University,\\ 
Department of Physics and Astronomy, 4242 TAMU, College Station, TX 77843, USA\\
$^6$Satellite Applications Catapult, Fermi Avenue, Harwell Oxford, Oxfordshire OX11 0QR, UK\\
$^7$NASA Goddard Space Flight Center, Code 660, MD 20771, USA\\    
$^8$Space Science \& Engineering Division, Southwest Research Institute, 
    P.O. Drawer 28510, San Antonio, TX 78228-0510, USA\\
$^9$IASF Palermo, Via Ugo La Malfa 153, 90146 Palermo, Italy. \\
$^{10}$The University of Texas at San Antonio, 
   Physics \& Astronomy Department, 1 UTSA Circle, 
   San Antonio, TX 78249, USA.\\   
$^{11}$Instituto de Astrof'sica de Andaluc\'{i}a (IAA-CSIC), 
Glorieta de la Astronom\'{i}a s/n, E-18008, Granada, Spain.\\
}

\begin{document}
\label{firstpage}

\date{Accepted: 23 February 2015. Received: 19 February 2015. in original form 12 January 2015.}
\pagerange{\pageref{firstpage}--\pageref{lastpage}}

\pubyear{2014}
\maketitle


\begin{abstract}


We present the calibration of the \Sw \uvot grisms, of which there are two, providing 
low-resolution field spectroscopy in the ultraviolet and optical bands respectively. 
The \uvg covers the range \lam 1700-5000~\AA\  with 
a spectral resolution ($\lambda/\Delta\lambda$) of 75 
at \lam 2600~\AA\  for source magnitudes of $u$=10-16 mag, 
while the \visg covers the range \lam 2850-6600~\AA\  with a spectral 
resolution of 100 at \lam 4000~\AA\  for source magnitudes of $b$=12-17 mag. 
This calibration extends over all detector positions, 
for all modes used during operations. 
The wavelength  accuracy (1-sigma) is 9~\AA\  in 
the \ucG, 17~\AA\  in the \ung and 
22~\AA\  in the \visG.   
The range below \lam 2740~\AA\  in the \uvg and \lam 5200~\AA\  
in the \visg never suffers from overlapping by higher spectral orders. 
The flux calibration of the grisms includes a correction we developed 
for coincidence loss in the detector. 
The error in the coincidence loss correction is less than 20\%.  
The position of the spectrum on the detector only affects the effective area (sensitivity) 
by a few percent in the nominal modes, but varies substantially in the clocked modes. 
The error in 
the effective area is from 9\% in the \ucg to 15\%  in the \visc.
\end{abstract}


\setcounter{figure}{0}

\setcounter{table}{0}


\section{Introduction}


\label{intro}

The \Sw mission \citep{swift} 
was launched to provide rapid response to gamma-ray bursts (GRB) 
over the wavelength range from gamma-rays to optical with three 
instruments: 
the Burst Alert Telescope (BAT) to detect gamma-rays \citep{Barthelmy}, 
the X-Ray Telescope (XRT) to observe the X-rays \citep{Burrows}, 
and the Ultraviolet and Optical Telescope (UVOT) for UV-optical 
photometry and spectroscopy  \citep{Mason, Roming}.  
UVOT spectroscopy is enabled by the inclusion of two grisms, 
the \uvg (1700-5000~\AA) and the \visg (2850-6600~\AA). These are mounted in a 
filter wheel which also houses the \uv and visible lenticular filters. 

The \Sw grisms provide a window on the UV universe to complement the 
high resolution HST instruments with a rapid response, low resolution  
option for the community.  The {\it X-ray Multi-Mirror } (\xmm) 
Optical Monitor (\OM) \citep{Mason2001} grisms
provide a similar functionality but for somewhat brighter sources 
and without the rapid response. 
Earlier missions which provided \uv spectroscopy include the International Ultraviolet 
Explorer (\iue\footnote{ \changed The IUE wavelength (2 ranges 1150-2000\AA; 1900-3200\AA) 
resolution was 0.2\AA\  for high dispersion  and 6\AA\ for low dispersion.}) 
\citep{boggess}, and GALEX\footnote{ \changed The GALEX spectral resolution 
($\lambda/\Delta\lambda$) was 90 for the NUV band (1771-2831\AA) and 
200 for the FUV band (1344-1786\AA).} \citep{martin}. 

Since November 2008 the automated response sequence of the \Sw UVOT, which governs the early exposures 
after a BAT GRB trigger \citep{Roming}, has  included
a 50 second \uvg exposure provided the burst is bright enough in 
the gamma rays. 
So far, this has resulted in two well-exposed \uv spectra of 
GRB afterglows: for 
GRB081203A \citep{kuin08} and the bright  nearby GRB130427A \citep{maselli}. 
\Sw has also obtained spectra for many other objects. 
These include comets \citep{comets}, AGN \citep{Mehdipour}, supernovae, e.g., 
\citet{bufano09, Brown12} and recurrent novae \citep{byckling} where the rapid response 
of \Sw has resulted in unprecedented early multi-wavelength coverage.

\begin{table*}
   \begin{minipage}{145mm}
   \caption{UVOT Grism Specifications}
   \label{specifications}

   \begin{tabular}{@{}lccl}
   \hline
                                    &{\bf visible grism} &{\bf \uv grism} \\
   \hline
grating                             & \changed{300}        lines/mm            & \changed{500}          lines/mm\\
spectral resolution                 & 100    at \lam4000~\AA     & 75      at \lam2600~\AA  \\  
wavelength range (first order)      & 2850-6600  \AA                 &1700-5000     \AA \\ 
wavelength accuracy (first order)   & 44         \AA                 & 17 (35)~$^f$ \AA \\
no order overlap (first order)~$^a$ & 2850-5200  \AA                 & 1700-2740    \AA \\
effective magnitude range~$^b$      & 12-17      mag                 & 10-16        mag \\  
astrometric accuracy~$^c,^g$        & 4          \arcsec             & 3            \arcsec\\
scale                               & 0.58       \arcsec/pixel       & 0.58         \arcsec/pixel \\
dispersion (first order)            & 5.9        \AA/pixel at 4200\AA& 3.1          \AA/pixel at \lam2600\AA \\
flux above which 20$\%$ coincidence loss~$^d$ & {$10^{-14}$}\ {ergs\ cm$^{-2}$\ s$^{-1}$\ {\rm\AA}$^{-1}$}&
{$10^{-13}$}\  {ergs\ cm$^{-2}$\ s$^{-1}$\ {\rm\AA}$^{-1}$} \\
\zo b-magnitude zeropoint~$^e$      & 17.7        mag                 & 19.0        mag\\     
   \hline
   \end{tabular}

   \medskip

 {  \begin{tabular}{@{}l}
    $^a$\  in the \uv-grism, the range without 2$^{nd}$ order overlap depends on the 
        placement on the detector. \\
    $^b$\  limit depends on spectrum, see section \ref{maglimits}.\\
    $^c$\  of first order anchor point. \\
    $^d$\  for low backgrounds. \\
    $^e$\  in \uvg for 10\arcsec circular aperture centered on zeroth order after successfully correcting the astrometry.\\
    $^f$\  \unG.\\
    $^g$\  using {\tt uvotgraspcorr} in a crowded field.\\
 \end{tabular} 

 }
  \end{minipage}
\end{table*}

\label{overview}

The UVOT uses a modified Ritchey-Chr\'etien optical design where light from the telescope 
is directed towards one of two redundant detectors using a 45-degree mirror. 
A filter wheel allows selection of either a \uv or optical lenticular filter, 
a $white$/clear filter, a \uv grism, a visible grism, or a blocked position. 
Behind the filter wheel is an image 
intensifier configured to detect each photon event with a 2048x2048 pixel resolution.      

The \Sw UVOT grism filters are the flight spares for the \OM\  instrument. 
The grisms for both instruments were designed using a 
\zemax\footnote{http://www.zemax.com}\  optical model. 
The \Sw UVOT instrument design and build procedure was modified to avoid the 
molecular contamination which impaired the \OM\  \uv sensitivity. Therefore 
the sensitivity of the \uv grism is much better in the \uv than that of the \OM\  grisms.
The \uvot \visg optics were blazed at 3600\AA. 
However, the \uvot \uvg optics were not blazed;
therefore the second order spectrum of this grism is significant and has to be accounted 
for in the analysis where the orders overlap.

\begin{figure}
\includegraphics[width=88.0mm,angle=0]{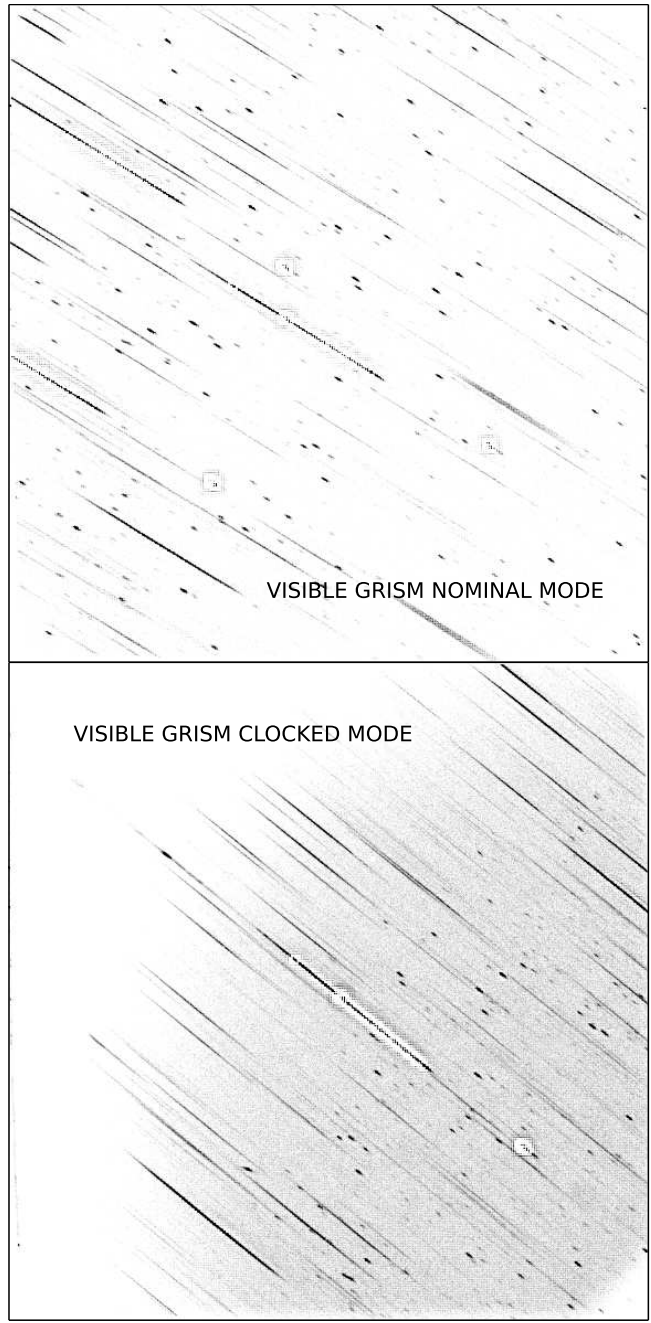}
\caption{A typical detector image of the \visG. The top panel 
was observed in nominal mode, the bottom panel in clocked mode. 
The zeroth orders are the short features, while the long lines are the first 
order spectra. In the clocked mode the zeroth order images are absent from the top-left of the image. 
Note also the change in the angle of the spectra 
on the detector between nominal and clocked mode. The nominal grism image has
a flat background, while in the clocked grism mode the background varies across the image. 
  }
\label{fig_3cd}
\end{figure}
%
Each of the two grisms can be operated in two modes. The so-called {\em nominal mode} 
is where the filter wheel is rotated so that the grism is positioned 
in direct alignment with the telescope optical light path.  However, in order to reduce 
the contamination by \zo emission of the background and field sources, in 
the so-called {\em clocked mode} the filter wheel is 
turned  so the grism is partially covered by the telescope exit aperture
which restricts the field of view, blocking some field stars and reducing the background 
light.
%
%
In the clocked mode the first order spectra of many stars in the field of view lie in the 
area uncontaminated by background or
the zeroth order spectra of other field stars as shown in Fig. \ref{fig_3cd}.  
The clocked mode has been used extensively, though the comet and GRB observations have been 
done in the nominal mode. 
A comparison of clocked and nominal images in Fig. \ref{fig_3cd} shows how effective 
the reduction of zeroth order contamination is from other sources in the field
for those spectra falling on the left hand side of the image.

\begin{figure*}
\includegraphics[width=135.0mm,angle=0]{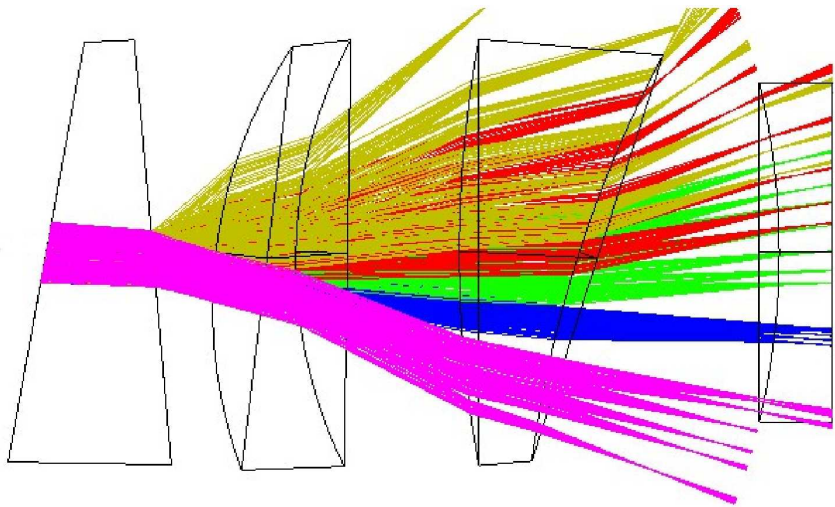}
\caption{Optical schematic of the \uvot \uv grism with rays for multiple wavelengths
and orders -1(pink), 0(blue), +1(green), +2(red), and +3(orange).  The computation 
was made for an on-axis beam coming in from the left, while it ends on the right 
at the entrance of the image intensifier. The green rays in the centre are of the
\uv-part of the first order. Long wavelengths of the first order can be seen interleaved
with red second and orange third order rays. 
        }
\label{fig_a}
\end{figure*}

The detector is a Microchannel plate (MCP) Intensified Charge coupled device (CCD) or MIC 
 \citep {Fordham89}. 
Each photon incident on the S20 multi-alkali photocathode can release an electron 
which is amplified a million-fold using a three stage MCP. The cloud of electrons 
hits a 
P46 fast-phosphor screen, producing photons
which are fed through a fibre taper to a CCD operated in frame transfer mode. 
The fibre taper reduces the footprint of the image intensifier output so it fits on the
exposed CCD area.
The exposed area corresponds to 256x256 CCD pixels, but after readout the photon splash is 
centroided to 8 times higher resolution, providing an effective image that is 2048x2048 
pixels square. 
The nature of the centroiding process is such that the effective size of each of the 8x8 
sub-pixels on the sky is not exactly the same, leaving 
a modulo-8 (\mod8) pattern in the untreated image which 
can be corrected for in data processing. 
However information loss that occurs when more than one photon splash is registered on 
a CCD pixel within the same CCD readout interval (coincidence loss) can cause some 
pattern to remain  for bright sources after correction. 

As noted above, the finite time over which each exposure is integrated on the CCD, the 
{\em frame time}, results in coincidence losses if the photon arrival rate is high 
enough  \citep{Fordham2000}.  
Statistically, there is a chance 
that multiple photons arrive within one CCD frame with spatially overlapping pulse profiles,  
in which case only one arrival will be recorded. This means that fewer source photons are 
detected than are incident on the detector, and the effect is larger when there is a higher
input photon rate,
resulting in a non-linear response with source brightness.  
Making use of the statistical nature of the effect, the coincidence loss can be 
corrected, and an expression for point sources has proved very effective 
in UVOT photometry, e.g.,  \citep{poole,breeveld10}. 
Extremely bright sources, above the brightness limit for coincidence 
loss correction,  suffer a further loss due 
to interference from events registered in neighbouring CCD pixels.

The background due to dark current in the detector is very low; instead the sky 
background is the limiting factor for faint sources. 
The sky background in the grisms is comparable to that in the UVOT white (clear) filter 
since both grisms transmit the 2800-6800~\AA\ optical band.

The sensitivity of the \uvot lenticular filter exposures is
decreasing slowly with time. 
In the \uv-filters the loss is about 1\% per year \citep{breeveld11}, while 
in the $v$-band optical filter it is larger, 1.5\% per 
year\footnote{Updates to the UVOT calibration documents are at 
http://heasarc.gsfc.nasa.gov/docs/heasarc/caldb/swift/}. 
Most likely ageing of the MCPs (proportional to lifetime photon throughput) is the main cause of the decreasing sensitivity 
which will affect all filters equally, while aging of the filter itself 
explains the different rate in the $v$-filter.
The 1\% sensitivity decrease is assumed to apply when the grism is employed
as well as the lenticular filters and is used for the grism 
calibration and data reduction.

The grism image is usually stored as an accumulated image on board the spacecraft, 
although it is also possible to record the data as an event list of photon times and positions. 
After transmission to the ground the data is routinely processed into raw images, 
which are corrected for the \mod8 pattern, followed by a correction 
due to small distortions in the fibre taper\footnote{the fibre taper distortion 
correction was determined using the lenticular filters} into a {\it detector image}.
The detector image is the basis for the spectral extraction. 

Knowledge of the position of the spectrum on the detector is 
crucial for determining the wavelength scale.
The \uvot spectra are formed by slitless dispersion
such that the  detector position
depends on the position of the source on the sky.
To define 
the position of the spectrum on the detector we use 
the position of a particular wavelength in the first order, 
referred to as the {\it anchor position}. The anchor position 
in the \visg is found at 4200\AA, in the \uvg at 2600\AA.

Table \ref{specifications} provides a summary of the capabilities of the grisms. 
More details can be found in the main body of the paper. 

The rest of this paper is organised as follows: 
first we provide more details on the \uvot grisms and the optical model,
and discuss the appearance of the grism spectrum on the image.   
We discuss how we map the sky position to that of the 
anchor point of the spectrum on the image.
Next we discuss the calibration of the dispersion. Then we present our correction for
the coincidence loss in the spectra and we determine the effective areas 
by including the coincidence loss corrections. 
After discussing the second 
order effective areas, we consider the zeroth order effective area and 
derive a photometric method and zeropoint for the \uvG.  
Next follows a description of the method used for extracting the spectra, 
and the related software. We conclude by describing how to use the 
\uvot grism, and some additional information that can be useful for the 
user.

In the following, we use the term {\em default position} for the spectrum 
placement in the middle of the detector, roughly at the boresight of the 
instrument. This is the normal operating position for sources; other positions 
are at an {\em offset}. The term {\em anchor} is used to designate the position 
that fixes the wavelength scale, in that the dispersion is measured relative 
to that position. By {\em model} we mean the \zemax\  optical model, either corrected 
or not.


\section{Description of the grisms and the optical model} 
\label{model}

The {``{grisms}"}  are actually made up of two optical elements: a grism and a tilt 
compensator as shown in Fig. \ref{fig_a}. 
The grism provides on-axis dispersion by means of a prism and transmission 
grating, but the focal plane is tilted. The tilt compensator element flattens the focal plane. 
The \uvg is direct ruled on a {\em Suprasil} substrate, while the 
\visg is replicated {\em epoxy}. 
The grating in the \visg is blazed to maximise the transmission 
of the first order. 
The \uvg grating was not blazed. 
Because the optics are in the converging beam of the telescope, which 
uncompensated would induce a shift in focus, 
the leading surface is slightly convex. 
The dispersed light is refocussed before leaving the grism. 
The grism design was optimised for the \uvg in the first order around \lam2600~\AA, 
and in the \visg around \lam4200\AA.  

The \zemax\ model which was used to design the grism optics 
has been used to assist in the 
in-orbit grism calibration, but some adjustments were needed.  
A significant reason for adjustment is that   
the model does not include the fibre taper optics 
between the MCP and the CCD \citep{Roming}, and that
the glass catalogue of the model does not include  coefficients for the 
refractive elements below 2000~\AA.
The optical model  was modified for the clocked modes with an 
appropriate decenter and rotation around the optical axis of the 
grism assembly\footnote{The rotation is called a "tilt" in the Zemax model.}. 

The boresight of the model was aligned to the observed boresight for all modes, but then 
a further adjustment of (-60,0) pixels was needed for the \uvg to align 
the model drop in flux in the left top corner of the detector due to the clocking 
to match the observations from calibration spectra. 

\begin{figure}
\includegraphics[width=88.0mm,angle=0]{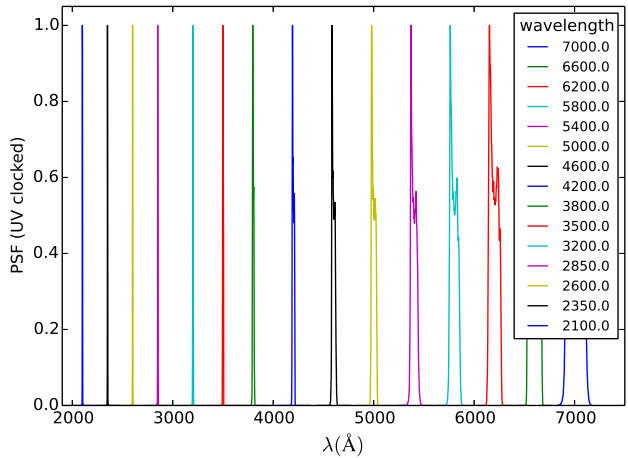}
\caption{The model point spread function of the \uvg is illustrated as a function of 
wavelength.}
\label{fig_b}
\end{figure}

The optical model predicts that the point spread function (PSF, which describes 
the distribution of where monochromatic photons fall on the detector) increases in 
size in the \uvg towards the red. 
This is illustrated in Fig. \ref{fig_b} where the 
2-dimensional model PSF has been integrated normal to the dispersion, 
illustrating the PSF variation as a function of wavelength.

The large width of the PSF in the \uvg for 
wavelengths longer than 4500~\AA\ causes the spectrum to appear smoothed out with wavelength  
at these wavelengths. 
In the UV, despite the small PSF width predicted by the model, other factors 
contribute to broadening the spectrum. 
Specifically, the actual PSF below 3000~\AA\  is thought to be dominated by the transverse spreading of 
the electrons leaving the cathode. Spacecraft jitter during the observation is removed 
using the shift and add method \citep{Roming}.
This results in a FWHM of the PSF of typically 3 pixels, about 10~\AA, in the 
\uv part of the spectrum in the \uvg and
about 20~\AA\  in the \visG.   

Fig. \ref{fig_a} displays a \uvg model calculation for an on-axis source of the 
rays for several \uv
wavelengths and for the five orders that can be registered on the detector. The first, second 
and third orders overlap, while the zeroth and minus-first orders are separate. 

The model zeroth order spectrum is dispersed in a hyperbolic fashion, as discussed in more detail
in Section \ref{zo_ea}.  
As a result photons with wavelengths longer than about \lam3500~\AA\ fall 
within a single pixel and in the \uvg the \uv spectrum forms a very weak 
tail that extends for 200 pixels, see for example the inset in Fig. \ref{fig_3a}. 

The dispersion angle, or slope of the spectra in the detector image, 
varies by about 5 degrees over the detector for a given mode. The 
angle near the centre of the image is different for each mode due to the positioning of 
the grism in the filter wheel and filter wheel clocking, and  is 
144.5$\degr$\ (\ucG), 151.4$\degr$\ (\unG), 140.5$\degr$\ (\visC), and 148.1$\degr$\ (\visN).

The model predicts the variation of the dispersion angle over the 
detector, but does not include the curvature of the spectra in the 
\uvG. The predicted model dispersion angle is used in the spectral 
extraction; see Section \ref{extraction}.

%
\section{The appearance of the spectrum on the detector}

The appearance of the grism spectrum differs between the \uv and visible grisms.
This applies equally to nominal and clocked modes in the same grism. 

In both grisms, the zeroth order extends over several pixels, and the higher 
orders overlap. However, in the \visg the zeroth order brightness and 
second order brightness are much less than the first order. As a result, 
the contamination of the first order spectrum from higher order light 
is small, and contamination by zeroth orders of field stars is usually 
small. The spectral tracks in the visible grism are straight, which 
means the spectrum is easy to extract. The angle of the spectra
in the detector frame varies slightly over the detector, however. Example images 
are shown in Fig.~\ref{fig_3cd}.
   
\begin{figure}
\includegraphics[width=88.0mm,angle=0]{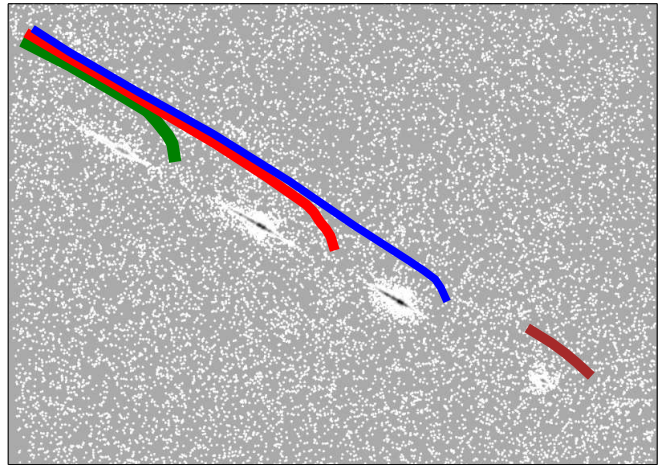}
\caption{A section of a ground calibration image of a narrow band filter exposure 
around 260~$nm$ in the \uvg outlines the positions of the orders. 
A cartoon of the typical relative positions of the spectral orders is shown 
above the data. The 
first order in blue, the second order in red, the third order in green, 
and the zeroth order in brown. The curvature of the spectral orders, which only 
occurs in the \uvg has been exaggerated for illustration.   }
\label{fig_2b}
\end{figure}

The zeroth and second order in the \uvg are of comparable brightness 
to the first order. Like in the \visg the zeroth order is extended, with a 
very small tail due to the  \uv response, and the first and higher orders 
tend to overlap. Unlike the \visG\  the \uv part of the spectrum in each order is 
generally curved. 
The curvature is largest in the upper right and lower left 
corners of the detector image and goes in the opposite direction. 
The spectral track is straight near the centre
of the image, where there is full overlap of first and second orders. 
Depending on where the spectrum 
falls on the detector, the second order overlap can start in the first 
order as soon as 2740\AA\ or as late as 4500\AA.   
A simplified drawing of the 
layout of the \uvg image can be found in Fig.~\ref{fig_2b}. 

%

\label{curvature}

The appearance of the observed curvature and displacement in the \uvg 
is evident in images with a very bright source, e.g., Fig. \ref{secondorder1}.  
The magnitude of the effect is 
a function of detector position, being very small at the default position.  
Adopting a straight line in the dispersion direction as reference, 
the maximum curvature offset in the cross-dispersion direction
varies from about 16 pixels in the lower left up to minus 25 pixels  
in the top right hand detector corner.

In the \visg there is no noticeable curvature or any offset of 
the higher orders at any point on the detector.

\subsection {Bright sources} 

\begin{figure}
\includegraphics[width=88.0mm,angle=0]{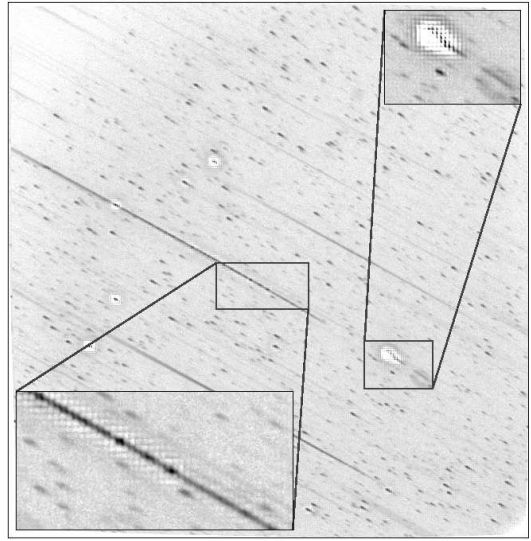}
\caption{A typical detector image of the \uvg in nominal mode.
The top inset highlights  
the extended disconnected \uv-tail in the zeroth order of the 
UV-bright source in the centre. The disconnection is due to a combination of 
detector effective area and interstellar extinction. 
The background around the bright zeroth 
order has been eaten away due to coincidence loss, and shows up as white pixels. 
The bottom inset shows the \mod8 pattern around a bright spectrum. 
Notice the many zeroth orders from weaker sources. }
\label{fig_3a}
\end{figure}

One of the characteristics of the grism images is that there is a \mod8 pattern 
of dark (low count rate) pixels next to the spectra of bright
sources; 
see for example the nominal mode \uvg 
image of the region around WR52 in Fig. \ref{fig_3a}, with the highlighted 
WR52 spectrum in the centre.  The cross-hatched \mod8 pattern is a sign 
that coincidence loss is present in the spectrum. 
Even when it is present, a correction for the coincidence loss 
is often still possible.  

For sources with a $b$ magnitude brighter than about 17  
the zeroth order in the \uvg develops a dark patterned region 
because of coincidence loss, and when brighter 
than 13$^{th}$ magnitudes, a region with a 49 pixel radius around 
the source is affected and can cause part of a nearby 
spectrum to be unreliable.

Occasionally, very bright stars ($V < 8^{th} $~mag) are in a grism image. 
These can cause problems because their readout streak 
(caused by exposure of the column during the image transfer to the CCD readout area)
leads to columns of elevated counts across the image, e.g., \citet{page13}. 
When the readout streak crosses a spectrum, it does so 
at an angle and background subtraction and the correction for 
coincidence loss may be affected. 
 
In parts of calibration spectra brighter than the coincidence-loss 
upper limit of 5 counts per frame 
a smaller count rate is observed than the expected count rate of nearly 1. 
This is thought to be caused 
by the amplified photon splash saturating neighbouring pixels.  
This would cause the centroiding to fail in which case those events are not counted.  
However, at such brightness the coincidence loss distorts the source spectrum  severely 
so confusion with the spectrum of a fainter source is unlikely.  

Other features due to scattered light that are seen in the lenticular 
filter images  \citep{breeveld10} may also be present in grism images. 
Experience shows that they do not occur frequently enough to affect 
the grism spectroscopy in practice.

%
\subsection{The fibre taper distortion}

\label{grism_distortion}

Images taken through the UVOT lenticular filters are spatially distorted by the image 
intensifier and fibre taper, with a small contribution from the lenticular filter itself.  
The correction has been determined to be the same for 
all lenticular filters, 
so a single distortion correction is needed to map positions on the detector to 
those on the sky. 
Using this distortion map, 
the standard ground processing produces a corrected image called the
{\em detector image}.  

The grism causes further distortion and this might be wavelength and order dependant. 

In order to do the anchor point calibration, we need to 
correct for the distortion of the anchor points, so we can find 
a mapping from the sky position to the anchor position.
We also need to map the zeroth order positions  
so we can get an astrometric solution for the grism image. 

The same distortion map as used for the lenticular filters is used to 
convert the raw grism image to a detector image.  It takes out the major 
distortion due to the image intensifier and fibre taper, though it 
may over-correct somewhat since it also includes the lenticular filter 
part of the correction. 

The model predicts the grism distortion, so it theoretically 
could be used to do the mapping from sky to anchor position 
after correcting for the image scaling caused by the fibre taper.  
However the model does not predict this distortion completely accurately.
Some differences remain from the observed anchor positions 
which are considered likely to be due to the unknown overcorrection 
due to the lenticular filters.    

The distortion in the detector image of the zeroth order 
positions  due to the grism optics 
was calibrated using catalog positions of the {\tt USNO-B1} 
catalog for several fields, and is made 
available\footnote{swugrdist20041120v001.fits} in 
the \CALDB\footnote{http://heasarc.gsfc.nasa.gov/docs/heasarc/caldb}.

%
\begin{figure}
\includegraphics[width=88.0mm,angle=0]{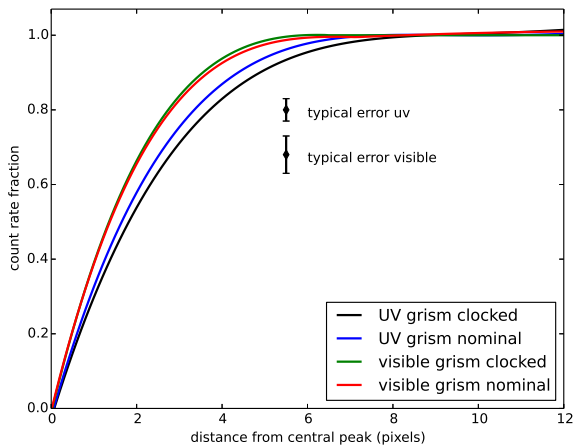}
\caption{The cross-dispersion count rates within a slit as function of the pixel
distance to the centre of the spectral track level off around 7.5 pixels in the 
\uvG, and around 5.5 pixels in the \visG. The profiles shown are for weak 
spectra with the lowest possible coincidence loss (WD1657+343).}
\label{fig_e}
\end{figure}


\subsection{The cross-dispersion profile}
\label{x-profile}

The final footprint of the light entering the detector is primarily broadened  
by transverse diffusion of the electron cloud in the gap between the cathode and MCP, which has a 
profile similar to a Gaussian. 
The grism optics also add broadening so that the spectral 
profile is of different width for the different orders.  
Finally, coincidence loss further affects the profile. 

In the \uvotpy\ (see Section \ref{softw}) a fit is made of the count rate as a function of 
pixel distance to the centre of the track using a gaussian distribution
\begin{equation}\label{sigma_eq}
f(x) = a \  e^{ - ({x-x_o \over \sigma})^2 }
\end{equation}
with $x$, the cross-dispersion pixel coordinate, $a$ the peak count rate, 
$x_o$ the centre of the spectral track, and $\sigma$ controls the width 
of the gaussian fit. 

In the first order \uvG, $\sigma$ is about 2.9 pixels at 1700~\AA, 
growing to about 3.3 pixels at 6000\AA. The second order is broader, 
with $\sigma \approx 4.5 $ pixels wide. The values are slightly smaller 
for the \visG, at 2.7 pixels.  

Measurements were made of the profile normal to the dispersion, 
which we will call the {\em cross-track} profile. 
The measurements were made in the region of no order overlap 
by repeatedly extracting the spectra with varying extraction widths.
The cross-track profile is not completely gaussian, but  
falls off more steeply in the wings. 
The plot of the enclosed count rates, similar to encircled energy in 
a point source but for a linear feature (see Fig. \ref{fig_e}) 
shows the cumulative distribution starting from the centre
as a function of the pixel distance from the centre to the 
border of the extraction ``slit".  
The width of the spectral track is seen to be smaller for the \visg than
for the \uvG. 
This profile can be used as an aperture correction, see Section \ref{aperture_correction}. 

The width of the spectral track changes with 
increasing coincidence loss due to the developing \mod8 pattern. 
This variation introduces an uncertainty larger than 20\% in the 
aperture correction when the coincidence loss is more than 20\%.
Therefore, a smaller aperture for the spectral extraction
with aperture corrections should only be used for faint spectra.  


\section{Calibration approach }


Before a full calibration of the wavelength and flux 
could be attempted, we needed to 
have a good aspect correction for the grism images so we know the sky location
of the boresight, and measure the width and curvature 
of the spectra over the detector. 
These basic calibrations were done first.

We have adopted an approach which merges the  calibration 
observations with the grism \zemax\  optical model. 
The optical model uses the optical set-up to predict the 
dispersion, order overlap, PSF, and throughput, 
and predicts their variation as a function of the source position 
in the field of view.  
The model contains a major part of the physics of the optics 
and thus it constrains the calibration, provides a way to verify 
observed parameters, and allows us to extend the calibration to all parts 
of the detector.  
As a result we can have a more reliable calibration by determining
corrections in the form of alignments and by scaling the model 
where appropriate. 

\subsection{Method of implementation}

For the \uv wavelength calibration sources, we selected 
Wolf-Rayet 
stars with a fairly good coverage of bright 
emission lines in the UV; 
see Table \ref{target_table}. 
These sources are on the upper range of brightness that can be observed with 
the UVOT grisms. 
For the calibration of effective area and coincidence loss we used mainly sources 
with reference spectra in  
the CALSPEC\footnote{http://www.stsci.edu/hst/observatory/cdbs/calspec.html}
database which are flux calibrated to typically 2-3\%.  

We selected anchor positions for this calibration 
at 2600~\AA\  for the \uvg and 4200~\AA\  for the \visg in the {\em first} order. 
Therefore, the anchor position on the detector is similar to the position of the source 
in the raw image in a lenticular filter when taken alongside the grism exposure. 


The curvature was measured relative to the model dispersion angle for the first and 
second orders for all \uvg calibration spectra and a correction was derived, expressed in 
terms of polynomials. The polynomial coefficients vary with the anchor position
of the spectrum but the curvature is always the same in the same grism mode at the 
same detector position. Bisplines were fitted to the polynomial coefficients, 
which allow the retrieval of the curvature at any anchor point on the detector.
The coefficients of the curvature calibration have been implemented in the 
\uvotpy\  \citep{uvotpy}, see Section \ref{softw}.
The spectra in the \visg are not measurably curved, but straight. 


The variation of the width of the first and (where possible) second 
spectral order was determined in the \uvg and also compared from image to image. 
Because of width variations along the spectrum, 
the spectral extraction was designed 
to keep the same enclosed energy for a consistent flux calibration 
by adjusting for slow variations in the width of the spectrum during spectral extraction. 

Once we understood broadly the geometry,
for each grism mode the 
analysis of anchor position, dispersion, coincidence loss and effective area 
was repeated as discussed in the next sections.

\begin{figure}
\includegraphics[width=88.0mm,angle=0]{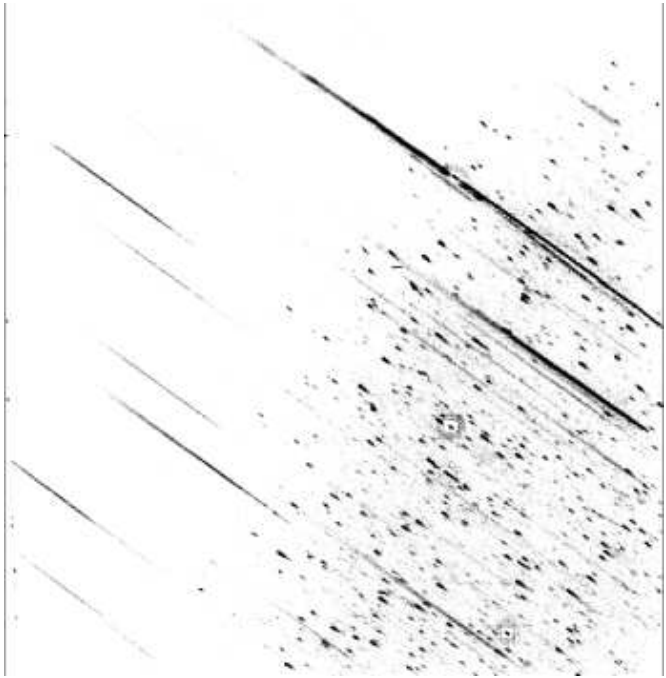}
\caption{A typical detector image of the \uvg in clocked mode.}
\label{fig_3b}
\end{figure}


\section{Calibration of the anchor point}


\subsection{The astrometric correction of the grism image}
\label{astrometry_correction}

The aspect is initially corrected 
using the best attitude from the spacecraft which is accurate typically 
to within 1.3\arcsec ($1\sigma$); see \citet{breeveld10}.

The positions of weak zeroth orders can be used to derive an improved 
aspect solution. 
The aspect correction is done by applying the {\tt uvotgraspcorr} 
\ftool\footnote{http://heasarc.gsfc.nasa.gov/docs/software/ftools/}
which uses the appropriate distortion file from the \CALDB\ 
described previously in Section \ref{grism_distortion}. 
The aspect corrected coordinates are written to the \fits header 
as the {\tt WCS-S} world coordinate system keywords including the keywords 
for the Simple Imaging Polynomial (SIP) convention \citep{SIP}
which capture the zeroth order anchor distortion.

\begin{figure}
\includegraphics[width=88.0mm,angle=0]{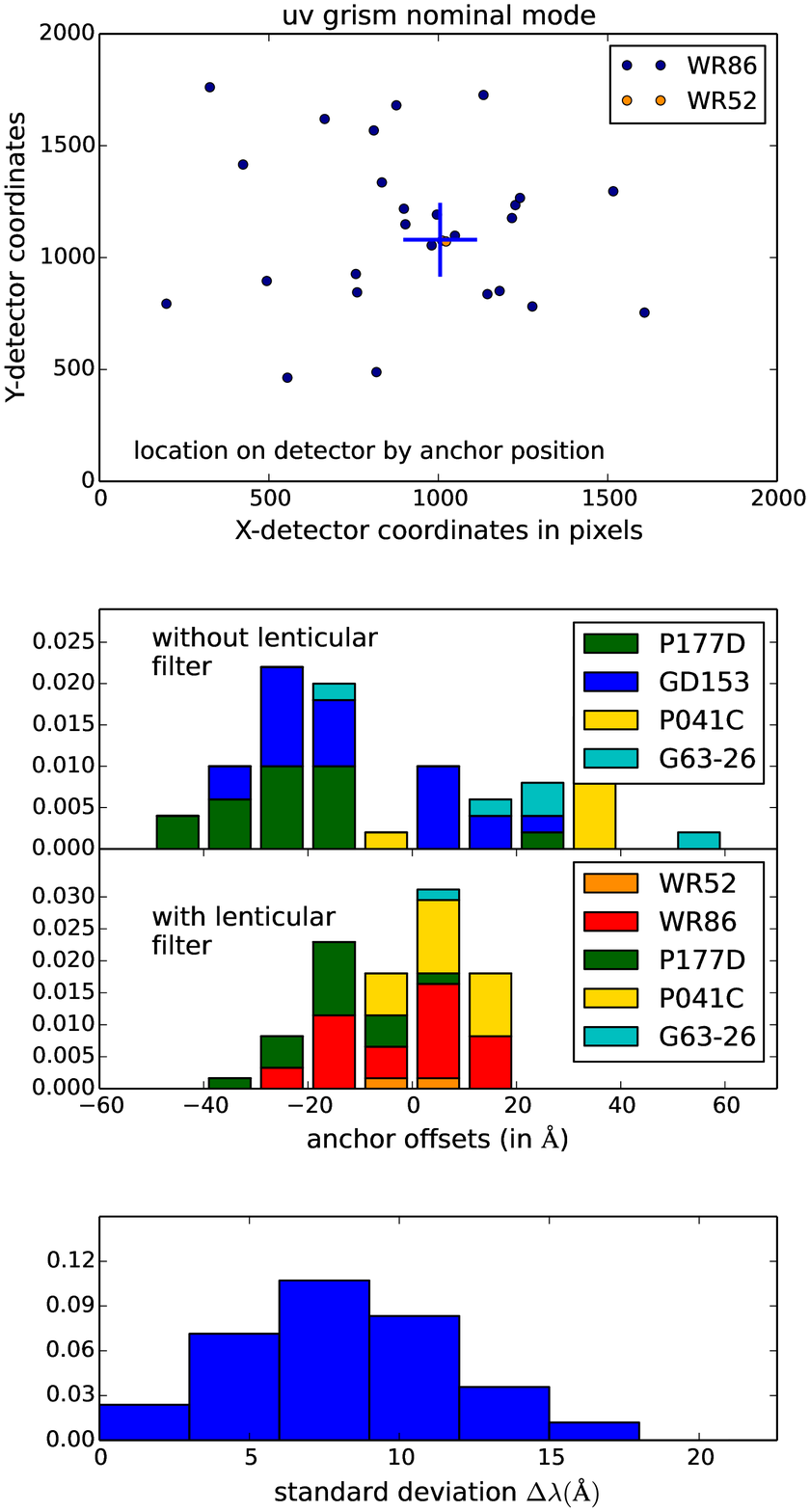}
\caption{The anchor and wavelength calibration for the \unG. 
The top panel shows the positions  on the detector 
for each spectrum used in the wavelength calibration. 
The position of the anchor for a 
spectrum at boresight is indicated with a blue cross. 
The second panel shows the measurement of 
the wavelength shift due to errors in the anchor position for two 
methods, by using {\tt uvotgraspcorr}, and with a lenticular filter 
alternatively.
The third panel shows the histogram of the standard deviation of 
the errors in the measured 
wavelengths after removing the anchor error. 
 }
\label{accuracy_uvnom}
\end{figure}

The accuracy of the astrometric correction depends on the success of the 
{\tt uvotgraspcorr} program. In the visible grism, with weaker 
zeroth orders due to the blazing, the errors tend to be  
larger. These are reported in terms of the anchor point accuracy in 
Table \ref{default_anchor}, and in the middle panel of Figs. 
\ref{accuracy_uvnom}, \ref{accuracy_uvc}, \ref{accuracy_visclocked}, and 
\ref{accuracy_visnominal}. 
The results show a scatter that varies mainly by target while for a given  
target different images tend to have similar errors.   
Tests show that the program fails in 3-4\% of the fields, 
in which case the correction can still be done by hand.

  \begin{table*}
     \begin{minipage}{115mm}
     \caption{Calibration targets used.}
     \label{target_table}
     \begin{tabular}{@{}lrrrll}
         \hline
          name/ID  &sp.  & J2000 &position & used    & reference\\ 
                   &type & RA       & DEC &  for$^£$     & spectrum, notes  \\
         \hline	   
WR1        &WN4& 00:43:28.4 & +64:45:35.4 & 1 & IUE, *\\
WR4        &WC5+?& 02:41:11.7 & +56:43:49.7 & 1 & IUE\\
WR52       &WC4& 13:18:28.0 & -58:08:13.6 & 1 & IUE,\#\\
WR86       &WC7(+B0III-I)& 17:18:23.1 & -34:24:30.6 & 1 & IUE,\#\\
WR121      &WC9d& 18:44:13.2 & -03:47:57.8 & 2 & IUE, \$\\
WD0320-539 &DA& 03:22:14.8 & -53:45:16.5 & 3,4,5 & CALSPEC\\
WD1057+719 &DA1& 11:00:34.2 & +71:38:03.9 & 3,4,5 & CALSPEC\\
WD1657+343 &DA1& 16:58:51.1 & +34:18:53.5 & 3,4,5 & CALSPEC\\
GD153      &DA1& 12:57:02.3 & +22:01:52.7 & 5 & CALSPEC \\
GSPC P177-D&F0V& 15:59:13.6 & +47:36:41.9 & 3,4,5 & CALSPEC\\
GSPC P 41-C&F0V& 14:51:58.0 & +71:43:17.4 & 3,4,5 & CALSPEC\\
BPM16274   &DA  & 00:50:03.7 & -52:08:15.6 & 4,5 & ESO HST standards\\ 
GD108      &sdB & 10:00:47.3 & -07:33:31.0 & 4,5 & CALSPEC\\ 
GD50       &DA2 & 03:48:50.2 & -00:58:32.0 & 4,5 & CALSPEC\\ 
LTT9491    &DB3 & 23:19:35.4 & -17:05:28.5 & 4,5 & CALSPEC\\ 
WD1121+145 &sdB & 11:24:15.9 & +14:13:49.0   & 3,4,5 & CALSPEC\\ 
G63-26&sdF & 13:24:30.6 & +20:27:22.1 & 3,4,5 & STIS-NGSLv2\\ 
AGK+81 266 &DB2& 09:21:19.2 & +81:43:27.6 & 5 & CALSPEC\\ 
BD+25 4655 &DB0&15:51:59.9  & +32:56:54.3 & 5 & CALSPEC\\
BD+33 2642 &B2 IVp & 15:51:59.9 & +32:56:54.3 & 5 & CALSPEC\\
         \hline
      \end{tabular}
   \medskip
 {  \begin{tabular}{@{}ll}
 use:$£$ &1: \uv grism anchor and wavelength calibration \\
      &2: visible grism anchor and wavelength calibration \\
      &3: \uv grism flux calibration\\
      &4: visible grism flux calibration\\
      &5: coincidence loss calibration\\
  CALSPEC &The Hubble Space Telescope calibration spectra data base at STScI \\
  IUE & ESA Vilspa archive for the International Ultraviolet Explorer  \\
  $*$ & \citet{HamannWN} \\
  $\$$ & \citet{TorresAtlas}, used WR103 to ID lines \\
  $\#$ & $CDS$ catalog III/143 \citet{Torres} \\
  & spectral types from \citet{vdhucht}, \citet{cookesion}, CALSPEC.\\
  STIS-NGSLv2& http://archive.stsci.edu/prepds/stisngsl/ \\
  ESO HST standards & http://www.eso.org/sci/observing/tools/standards/spectra/hststandards.html \\
    \end{tabular} 
  }
   \end{minipage}  
\end{table*}

\subsection{Reference data}
\label{wavecal}
The anchor and wavelength dispersion calibration consists of the determination of
the scaling of the model by using calibrated spectra of bright Wolf-Rayet (WR) stars. 

Our WR stars were observed by IUE, and have also 
ground based spectra available with sufficiently good wavelength 
calibration to determine the wavelengths of spectral emission lines. 
Major lines used in the WC-type spectra are: 
Si~II 1816~\AA, C~III 1909~\AA, 2297~\AA, 
C~IV 2405, 2530, 2906~\AA, 
O~IV 3409~\AA, C~III 4069, 4649~\AA, 
C~IV 5801~\AA, while for the WN-type spectra we used:
He~II 2511, 2733, 3203, 4686~\AA.
Further minor emission lines, sometimes blended, are present in our spectra.  
For  WR121, which was used for the \visG, no IUE spectrum is 
available, but for the same spectral type 
the IUE spectrum of WR103 provides a similar spectrum 
with the line identifications from \citet{Niedzielski} 
and a good ground based spectrum from \citet{TorresAtlas}. 
For WR4 no ground based spectrum was used, but the lines were easily 
identified from comparison to the other spectra. 
Though WR86 is a binary, the spectrum is dominated by the WC spectrum;
radial velocities may lead to shifts of $<2$~\AA\   which are negligible
at the resolution of the grisms.

For the wavelength verification some spectra from the flux calibration 
sources were used. The lines used in those spectra are mainly Mg II 2800\AA and
the Hydrogen lines.

\begin{figure}
\includegraphics[width=88.0mm,angle=0]{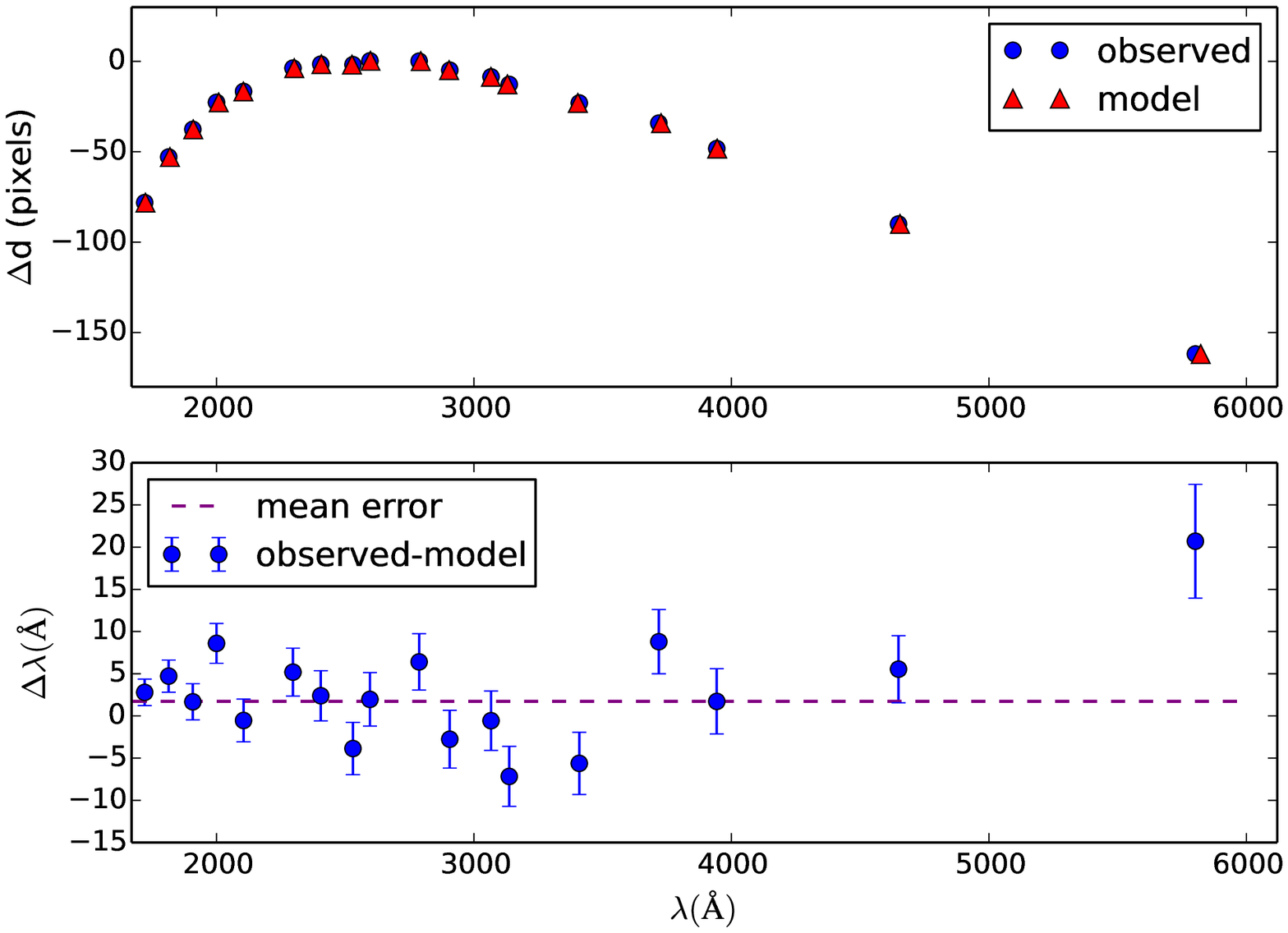}
\caption{Example of the determination of the wavelength accuracy for one 
calibration spectrum in the clocked \uv grism. In the upper panel
 $\Delta$d is derived from the anchor distance in pixels, after 
subtracting a linear constant dispersion factor, and illustrates 
the higher order variation of the dispersion. The lower panel shows the 
errors in the measured wavelengths.}
\label{accuracy_example1}
\end{figure}

\subsection {First order anchor position - fitting to model} 
\label{anchor1st}

In order to calibrate the anchor position, calibration spectra 
accompanied by an image in a lenticular filter 
were taken of WR stars, see Table \ref{target_table}, while the 
pointing was offset 
so that the spectra covered the detector; see the top panels in 
Figs. \ref{accuracy_uvnom}, \ref{accuracy_uvc},
\ref{accuracy_visclocked}, and \ref{accuracy_visnominal} 
where the locations of the anchor points on the detector are plotted.  
Although the emission lines in these stars are broad, their width is not an
impediment as it is close to the spectral resolution of the instrument. 

The calibration spectra were observed in a special mode, where a lenticular filter 
exposure is taken just before and/or after the grism exposure. 
The spacecraft pointing is not changed during the sequence although there may still be some drift in the pointing between the exposures.
Within an exposure, the positions are corrected using the on-board shift-and-add 
algorithm \citep{poole}. 
The position of the target in the lenticular images can then be correlated 
to that of the observed anchor in the grism.

\begin{table*}
  \begin{minipage}{145mm}
   \caption{Default anchor positions and wavelength accuracy.}
   \label{default_anchor}

   \begin{tabular}{@{}lcccccl}
   \hline
Grism mode& anchor$^{1}$ &anchor 2$\sigma$ &wavelength     &anchor 2$\sigma$ &wavelength\\
          & default      &accuracy(\AA)&accuracy(\AA)$^4$  &accuracy(\AA)&accuracy(\AA)$^4$\\ 
          & position     &detector centre$^{2}$ &detector centre$^{2,3}$ & full detector& full detector\\
   \hline
\multicolumn{6}{c}{anchor position determined using a mode combined with lenticular filter}\\
   \hline
\ungb&[1005.5,1079.7]   & 30 & 7,18,36 & 35 & 8,16,34  \\
\ucgb&[1129.1,1022.3]   & 12 & 8,11,21 & 17 & 7,22,18  \\
\visnb&[1046.3,1098.3]  & 30 & 5,10,6 & 44 & 6,13,6 \\ 
\viscb&[1140.7,1029.6]  & 48 & 5,14,13 & 44 & 4,13,12 \\
   \hline
   \multicolumn{6}{c}{anchor position determined using astrometry from {\tt uvotgraspcorr} }\\
   \hline
\ungb&[1005.5,1079.7]   & 53 & 46,15,22 &  53   & 51,17,25  \\
\ucgb&[1129.1,1022.3]   & 47 &  8,11,21 &  47   & 7,12,18  \\
\visnb&[1046.3,1098.3]  & 88 &  3,10,8  &  88   & 5,13,7 \\ 
\viscb&[1140.7,1029.6]  &118 &  9,16,14 & 118   &  8,16,12 \\
   \hline
   \end{tabular}
   \begin{tabular}{@{}ll}
$1$\ &first order, in detector coordinates  \\  
$2$\ &The detector centre is defined by image pixels between 500 and 1500 in X and Y.\\
$3$\ &2$\sigma$ errors for three ranges in the \uvg of 
$\lambda <2000\AA, 2000 < \lambda < 4500\AA,4500\AA < \lambda$,  \\  
   & and in the \visg of  $\lambda <3100\AA, 3100 < \lambda <5500\AA, 5500\AA < \lambda$. \\
$4$\ &excluding errors due to the anchor.\\     
   \end{tabular}
  \end{minipage}
\end{table*}

The position of the source relative to the boresight in the lenticular image and the 
anchor position relative to the boresight in the grism image are related in a 
fixed manner. 
Ignoring the distortion, the conversion from lenticular filter to grism position
is a shift in detector X,Y position and a scale factor. No rotation is neccessary, due to 
the coordinates being tied to the detector orientation. 

For each observed spectrum, bright spectral lines were identified in the image close 
to the anchor point, whereafter the anchor point on the detector 
image was determined for each spectrum.  
The anchor positions for the default position (with the source at the boresight) 
have been given in Table \ref{default_anchor} in {\em detector coordinates}\footnote
{The detector coordinates are converted here from mm to pixels by a centre of [1100.5,1100.5], 
and a scale factor of 0.009075 mm/pix}, and are shown in the top panel in Figs. 
\ref{accuracy_uvnom}, \ref{accuracy_uvc}, \ref{accuracy_visclocked}, and 
\ref{accuracy_visnominal} as a blue cross.

A comparison of the observed grism anchor positions and source positions in the 
corresponding lenticular filter(s) 
implies a pixel scale in the grism image 
of $0.58\pm 0.04$ arcsec/pixel, larger than the $0.502$ value
for the lenticular filter, though it should be noted that the 
pixel scale varies due to distortion.

In the lenticular image  the position of the source can be found from the sky position.
Given the source sky position and the \fits WCS header in the aspect 
corrected lenticular filter image, we derive the astrometrically 
corrected source position on the lenticular filter image and thus the 
source position relative to the boresight position. 
That relative position is converted into the field 
coordinate\footnote{the field coordinate is the angular coordinate relative to the boresight}
for input to the model. 
We now can use the  model  to find the anchor position on the grism 
detector image, provided the model has been properly scaled.  

We found that a simple scaling of the pixel size in the model brought 
all the model and observed anchor positions to within 16 pixels. 
The position differences were not displaced randomly, so a bispline 
fit to the X and Y coordinate differences provided a final correction. 
The order of the bisplines was kept 
as small as possible in order to keep the number of parameters 
low\footnote{e.g.,for the \ucg there were 24 parameters used to fit 
50 positional data; other grism modes used fewer parameters to fit
a comparable number of data points. Therefore, enough free parameters remain.}. 

The corrected model was used to tabulate 
a lookup table of anchor positions on a grid of field positions. 
The lookup table was used subsequently to rederive the anchor positions 
to obtain an estimate of the accuracy. 

Any inaccuracy of the anchor position leads to a shift in the wavelength 
scale.  
Since many calibration observations had a lenticular filter before 
and after the grism exposure, we also obtained an estimate of the pointing 
drift during the exposures (often 1 ks long) of typically 
6 pixels ($\approx$ 3\arcsec), though a larger excursion between exposures 
(10 pixels) was observed in a single observation. 
This is consistent with the accuracy seen in the anchor positioning.  
In the \ung calibration there was only one lenticular filter in the observation 
so no correction for drift between the grism and lenticular filter exposure
was possible. This explains the larger errors in the anchor position 
calibration (thus the wavelengths were shifted more). 

Calibration observations taken before 2008 did not include a lenticular filter. 
This includes most of the flux calibration spectra for the \visG.
When using the sky position from the image header to determine 
the anchor position without using a lenticular filter image taken next
to the grism image, the anchor error is due to the accuracy of the {\em uvotgraspcorr}
program as discussed in Section \ref{astrometry_correction} above.  

Verification with independent data was done using flux calibration targets 
with good enough spectral lines. The anchor offset was determined for the case 
that a lenticular filter was used for determining the anchor position, and also 
by using the {\tt uvotgraspcorr} program.  
The wavelength calibration sources were not considered, since they were already  
used to determine the mapping from the first order boresight to the zeroth 
order boresight position (used in the grism WCS-S). 
The middle panels in Figs. \ref{accuracy_uvnom}, 
\ref{accuracy_uvc}, \ref{accuracy_visclocked}, and 
\ref{accuracy_visnominal} show histograms of the measured corrections to 
the wavelengths, taken at the anchor position.  The corrections derived 
for independent sources when using the lenticular filter method show a similar 
distribution as the WR sources used for the wavelength calibration. 
The corrections derived when not using a lenticular filter show a much larger spread. 
In particular for the visible grism an apparent bias in the offset appears. 
This is due to there being a large number of spectra taken under similar observing 
conditions for some sources which give a consistent offset for that single source. 
The derived accuracy is reported in Table \ref{default_anchor}. For the 
case of anchor position without a lenticular filter, the uncertainties are 
considered to be the same over the whole detector. 

\subsection {Second order anchor position} 
\label{secorderanchor}
The WR calibration sources have bright lines, and using the model as 
guidance, the much weaker second order \uv lines were identified in the 
\uvg calibration images. 
In the \visg the orders are weaker and overlap and this is not possible. There the 
model prediction was used. 

Similar to the first order we define an anchor position in the second order 
at a fixed wavelength. For the calibration we determine the distance 
of the second order anchor to that of the first order.
Errors in the first order anchor position are easier to correct that way than 
by working directly with the second order position on the image.  

Using the observable lines in the \uvG, an anchor position of  \lam2600~\AA\ in second 
order was determined by interpolation, and sometimes extrapolation,
as well as a scaling factor for the second order spectral dispersion for
the areas on the detector where there was not much overlap of first and
second order. 
  
The distance of the first and second order anchor was thus derived.
No scaling of the model was attempted on the first to second order distances 
which were fitted with a bispline instead. 
By interpolation we obtain a prediction of the anchor position  
for the area of complete order overlap. 
The main source of error in the second order parameters 
is  the error in the first order anchor position. 


\section{The dispersion of the spectra} 


\begin{figure}
\includegraphics[width=88.0mm,angle=0]{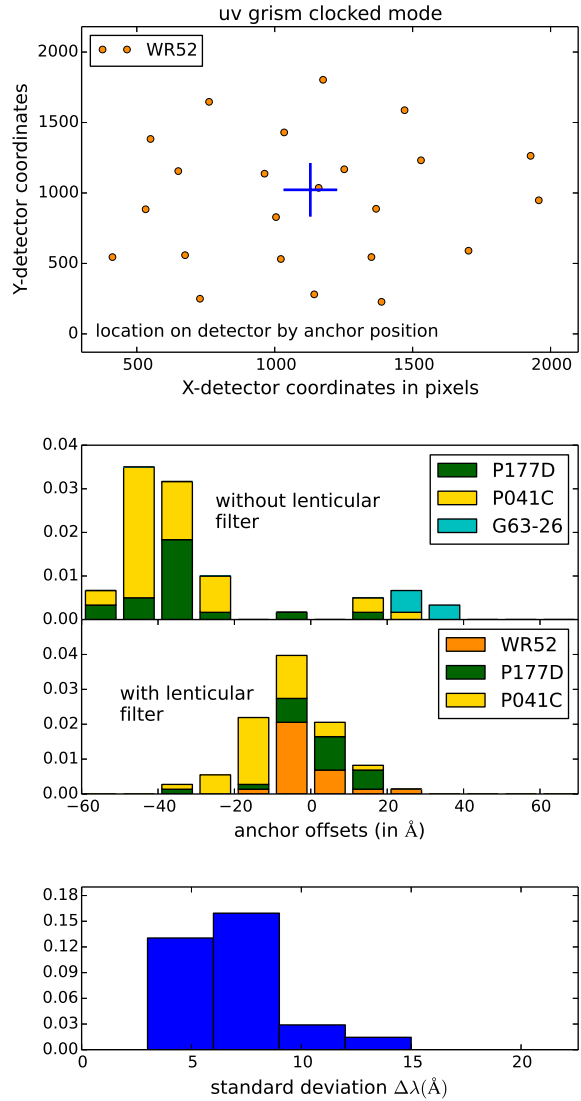}
\caption{Accuracy of the wavelength calibration  for the \ucG. 
See Fig. \ref{accuracy_uvnom} for explanation of each panel. }
\label{accuracy_uvc}
\end{figure}

\begin{figure}
\includegraphics[width=88.0mm,angle=0]{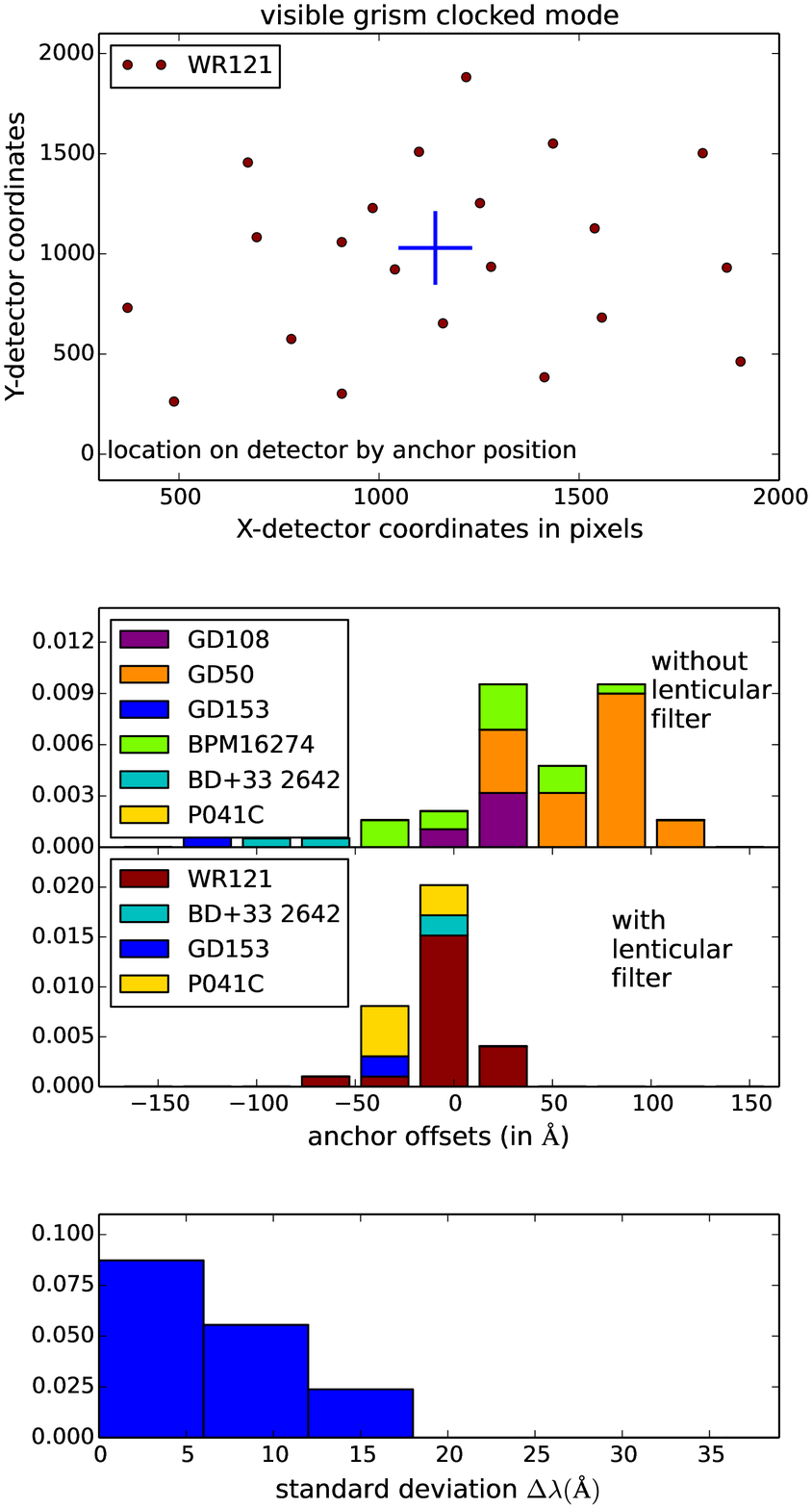}
\caption{Accuracy of the wavelength calibration for the \visC, see 
Fig. \ref {accuracy_uvnom}. See Fig. \ref{accuracy_uvnom} for 
explanation of each panel.  }
\label{accuracy_visclocked}
\end{figure}
\begin{figure}
\includegraphics[width=88.0mm,angle=0]{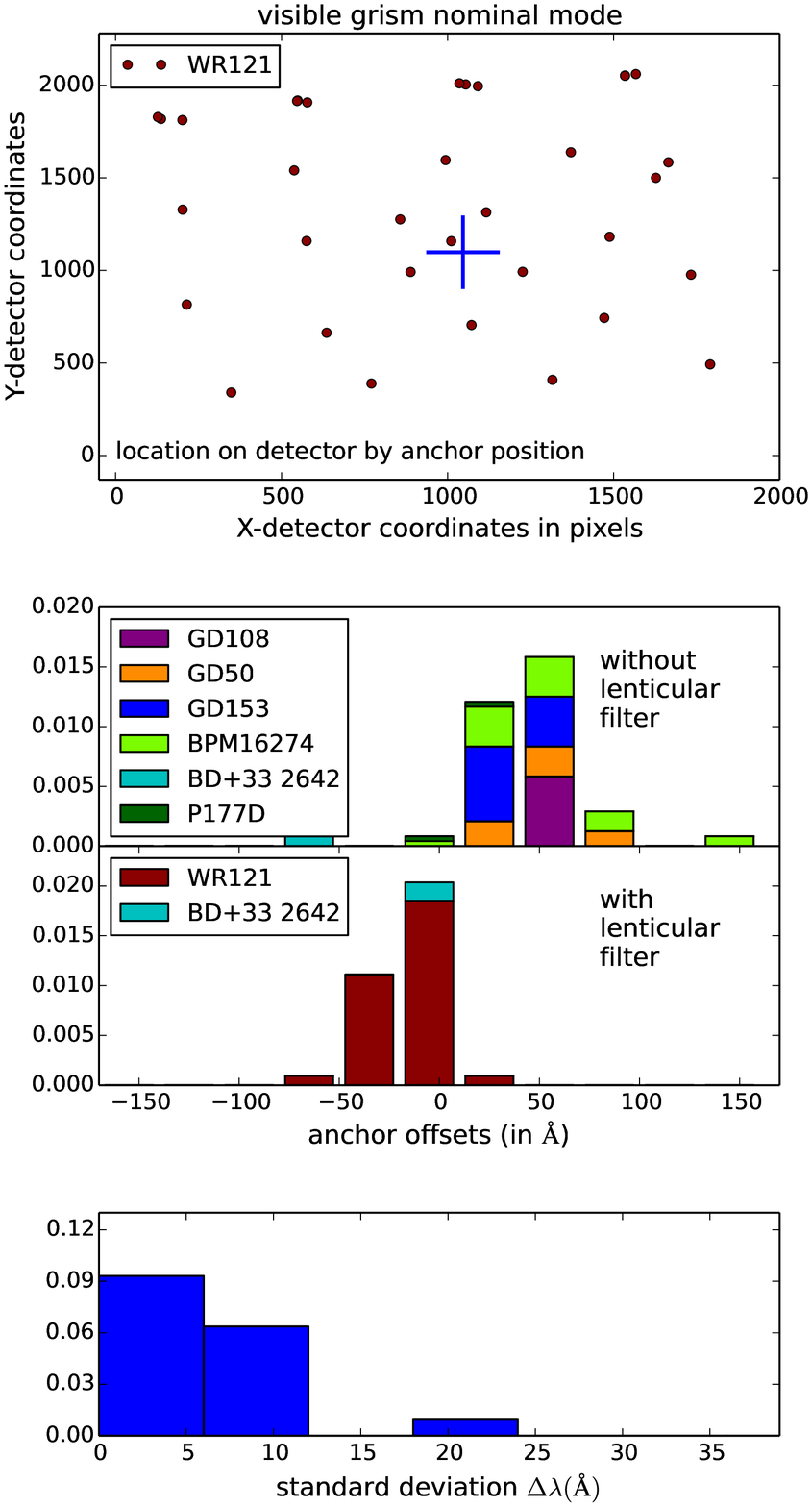}
\caption{Accuracy of the wavelength calibration for the \visN, see 
Fig. \ref {accuracy_uvnom}. See Fig. \ref{accuracy_uvnom} for 
explanation of each panel.}
\label{accuracy_visnominal}
\end{figure}

\subsection{The method used}

For the anchor position calibration, the anchor in each spectrum had already been determined 
by examining the position of the nearby lines on the image. We could consider the 
positions of the other lines in the image relative to that anchor and compare to the 
predicted positions from the \zemax\ model. 

The comparison showed that a scaling had to be applied to the model to properly 
fit the observed spectral line positions in the calibration spectra.  
For each spectrum the pixel position of lines relative 
to the anchor position were measured, while the wavelengths 
were taken from line identifications and/or the reference spectrum. 
The  measured wavelengths are affected by the \mod8 noise from coincidence loss because the 
calibration sources are very bright, which leads to a measurement 
error of about one pixel, except for wavelengths in the \uv grism above 4800A, where the 
error is slightly larger, see Fig. \ref{accuracy_example1}.  

The \zemax\ model calculations were made on a 28x28 grid of positions on the detector, 
referenced by the anchor position.  Each model spectrum was computed 
at a range of wavelengths. The same wavelengths were used in all spectra 
for the \uvG, and a second set was used for all spectra computed for the 
\visG.  
In order to interpolate the dispersion from the model, 
for the \uvg a fourth order and for the \visg a third order polynomial were fitted to the model.  
To determine the dispersion at any point on the detector, for each polynomial power, 
the polynomial coefficients of the nearest model grid points are interpolated 
using bilinear interpolation.  
Initially, this was used to determine a model dispersion at the actual position 
of the calibration spectrum.

The dispersion relation as determined from the Zemax model calculation was compared to the observed
line positions to refine the scaling factor needed for the model.  There is some spread 
in the difference between model and observed line positions. 
Using all lines in a spectrum, the RMS of the errors in the line positions 
is used as a measure of the accuracy of the dispersion. This was used iteratively to 
improve the scale factor used in the model. 

The resulting RMS error for all spectra, which cover the face of the detector by position in 
each grism and mode, 
have been shown as histograms in the bottom panels of Figs. \ref{accuracy_uvnom}, 
\ref{accuracy_uvc}, \ref{accuracy_visclocked}, \ref{accuracy_visnominal}.
The mean values of RMS are given in Table \ref{default_anchor}. 

The best fitting scale factor to the dispersion is close to a single value for the 
whole detector for each mode. However, some fine adjustments were neccessary. 
As noted in Section \ref{model}, the glass catalog of the model 
did not include  coefficients for the refractive elements below 2000~\AA, 
which were extrapolated. Also, the very extended nature of the PSF above
4500~\AA\ in the \uvg may lead to a different interpretation of the emission 
line peak location in the model and observed line positions.

The lowest order acceptable fits were adopted.  
In the dispersion calibration for the \ung only a constant was needed for 
an acceptable fit but in the \visn a bilinear fit was needed. 
In the \ucg a linear-quadratic fit, and in the 
\visc a bilinear fit were adopted. The \ucg also includes a linear dispersion 
scaling (i.e., it includes a wavelength dependent factor).  

Finally, using the scaled model obtained as described above, 
dispersion polynomial coefficients were 
tabulated for a 28x28 grid covering the detector. 
This allows for accurate interpolation at any anchor location. 
The table was written to the wavelength calibration file, one 
for each grism mode.

\subsection{Internal accuracy of the wavelength scale}

We next verified the dispersion. 
In order to verify the accuracy of the dispersion calibration file, 
we should have used independent observations. However, this would require
a prohibitive amount of extra observing time. We have used only 
a limited number of parameters to perform the model scaling and thus 
using the calibration observations (which have two to three times as many 
parameters) should give a valid indication of the accuracy of the calibration. 

While the calibration software accessed the \zemax\ model and applied the 
required scaling, for the verification we used the calibration file with  
the \uvotpy\ (see Section \ref{softw}).

As an example of  an individual spectrum, Fig. \ref{accuracy_example1} 
displays in two panels the wavelength accuracy. After taking out the linear 
term of the dispersion, the remaining variation in the dispersion shows small 
offsets between the values from the spectral extraction (observed) and those 
determined by careful calibration (model). 
The positions of spectral lines were taken from our calibration data and compared 
with the position predicted by the dispersion relation for that spectrum for 
those lines.  
This is presented in the bottom panel which shows as a function of wavelength the 
standard deviation of the 
difference $\Delta\lambda$ between predicted and measured wavelengths. 

The total wavelength error varies over the detector and is smallest near 
the default position.

\subsection{Wavelength calibration of the second order}

The second order in the \visg is weak, since the grism was blazed, and has not 
been considered for calibration.

\begin{figure}
\includegraphics[width=88.0mm,angle=0]{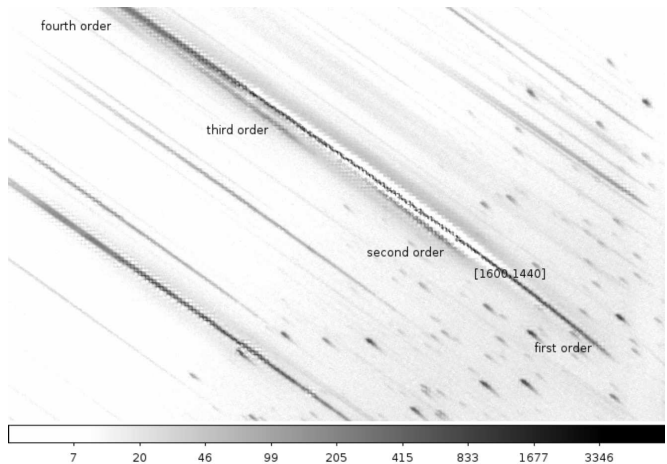}
\caption{Section of the \uvg image containing a very bright source spectrum in the 
top right corner.  Here the first order is overwhelmed by coincidence loss related 
\mod8 patterning and the higher orders clearly separated from the first order. 
The approximate anchor position is indicated as coordinates. }
\label{secondorder1}
\end{figure}

In the \uvg the second order separates from the first order most prominently 
in the upper right hand part of the detector. 
Fig. \ref{secondorder1} shows an annotated section of a grism image of a bright 
WR star. Compare also to Fig.~\ref{fig_2b} for a cartoon schematic.  
The [1600,1440] anchor position shows that the spectrum is in the upper right hand corner 
of the image. 

As mentioned in Section \ref{secorderanchor} 
the spectral lines in the WR calibration spectra were used. 
The  model predicts the second order, but 
the order distance and dispersion need to be scaled, just like for the first order.
The 1909~\AA\ line is usually seen clearly.  
In second order the stronger lines at 2297, 2405, and 2530~\AA\ fall close to the 
very strong 4650~\AA\  first order line, which complicates their identification. 
The 3409~\AA\ line is sometimes visible where the first order is tapering off, 
and adds a useful data point for the dispersion. 

This calibration was done for both \uvg modes, resulting in a solution for the 
second order dispersion. 
The predicted positions for the \ucg are 
within 50~\AA. The main reason for a large error seems to be that any error in the anchor 
position affects the second order scale nearly twice as much.

The first and second order overlap starts at different wavelengths depending 
on the position of the spectrum on the detector because of the curvature. 
Even at large offsets the 
second order and first order converge at longer wavelengths. 
That means that only the \uv part of the second order is useful. 
For the best case, at an anchor offset from the default position on the detector 
of $\Delta x > 2 \arcmin, \Delta y > 4 \arcmin$ 
the second order from 1700-2200~\AA\  will be unaffected by the first order, 
and will not 
contaminate the first order up to wavelengths of about 4500~\AA.

\begin{figure}
\includegraphics[width=88.0mm,angle=0]{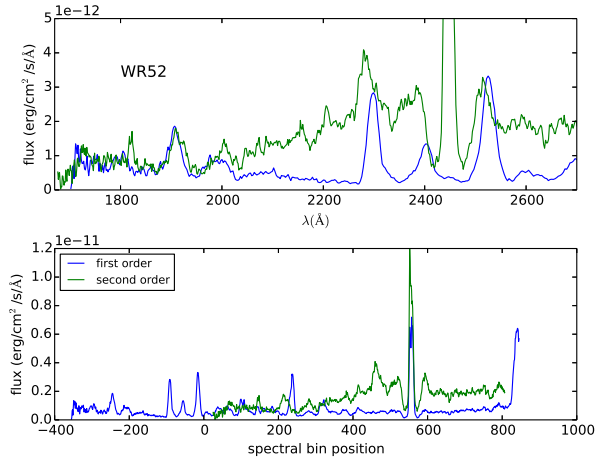}
\caption{First and second orders when the spectrum is at an offset from the 
default position (anchor position = [1480,1490]). 
The bottom panel plots the spectra as a function of 
position on the detector, while the top panel shows them as a function of wavelength.
The second order merges with the first order above 2000\AA\ in second order. 
In the bottom panel you can see that the 2nd order is influenced by the bright 
1st order 4649 line\AA. When plotted in wavelength it is not so obvious. The 
second order lines at 2297, 2405, and 2530\AA\ are bright enough to slightly affect 
the first order.  
 }
\label{firstandsecond}
\end{figure}

Fig. \ref{firstandsecond} shows the first and second orders for the bright star WR52.
The second order position relative to the first order was used when extracting the 
spectrum.  The second order is extracted using this position which partly overlaps 
the first order. Therefore, when the location of the second order extraction slit 
falls over the first order, the combined counts are obtained. 
Once this happens the plot shows a falsely brighter second order.
The same is done for the first order. In the region of overlap 
the effective areas of first and second order are different which accounts for 
the different derived flux in {``first''} and {``second''} order in the region of overlap.
It can also be seen that in the region of order overlap the bright 
second order \uv emission lines of 2297, 2405, and 2530\AA\ marginally affect the first 
order spectrum. The second order contamination is larger when the spectrum is brightest
in the \uv\ (see also Section \ref{EA2nd}).   

\section{Coincidence loss corrections}

\label{coincidence_loss}

In Section \ref{overview} we introduced the way the detector works, and 
the resulting potential for coincidence loss. 
Due to the finite number of frames in any measurement the statistics follow
a binomial distribution. 
Therefore the general form of coincidence loss for a 
point source can be determined as a function of the incident count 
rate and a correction factor can be applied. 

The geometry of the spectrum on the detector is different from that of 
a point source but, since the statistics are similar, the same approach can 
be taken. To account for the difference in geometry, the area on the 
detector for calculating coincidence loss has to be adjusted.

The in-orbit effect of coincidence loss on high backgrounds was investigated 
for background and point sources in \citet{breeveld10} who showed that the 
correction used for point sources was good for the typical backgrounds in 
\uvot photometry.
The spectra of the faintest sources have count rates that are smaller than 
the background rate. 
Therefore, we can safely assume that the coincidence loss correction 
formulated for the background will give the right correction for the 
faint spectra. This approach was used to derive the initial coincidence 
loss correction for \uvotpy\ version 1.


\subsection{The nature of coincidence loss in spectra}
\label{coi2}
As mentioned in the introduction, the finite frame time $t_f$ (typically 0.0110329 
seconds) of the UVOT MIC detector places limits on the ability to detect high count rates 
$C_r$ (in counts per second), since each detection area can only measure one count per 
frame. 
Let's define the counts per frame by the greek letter $\varphi$, which is then found from 
\begin{equation}
\label{define_phi}
\varphi = C_r t_f. 
\end{equation}
where $C_r$ and $t_f$ were introduced previously.
The count rate is computed using a certain area on the detector as discussed below.
The incident counts on the detector are assumed independent
so that an exposure of $N$ frames has a binomial distribution which in the limit
of a large number of frames turns into a Poisson distribution.  
Accounting for the fact that coincidences of multiple photons in a frame are only 
counted 
as one, the statistics then relate the measured counts per frame, defined as 
$\varphi_{m}$, to the incident counts per frame, defined as $\varphi_{in}$. 
For a simple point source the statistical relation is given by: 
\begin{equation}\label{basiccoi}
{\varphi}_{in} = {-1\over\alpha}\quad {ln(1 - \varphi_{m})},
\end{equation}
where $\alpha$ is the fraction of the  frame time when the frame is exposed, 
excluding the dead time for CCD frame transfer, and $ln$ is the natural logarithm. 
For the default \uvot full-frame readout $\alpha = 0.984$.
It has been assumed in this paper that the measured count rate has already been 
corrected for dead time, similar to the usage in \citet{poole}.

Coincidence loss tends to steepen the PSF in point sources, and \citet{poole} found 
that an aperture of radius $10.5\pm1.2$~pixels gave the optimal aperture range 
where photometry was the least affected by coincidence loss. Hence a circular 
aperture with a radius of 10~pixels
was chosen for the detection area for UVOT photometry in the lenticular filters. 
However, a circular area is not appropriate for the grism spectra. A different 
area for coincidence loss must be determined.

The idea of a coincidence area was proposed by \citet{Fordham2000} to explain 
that the coincidence loss they found for flat field illumination of the detector was
larger than that in a single pixel. Using an appropriate coincidence area for determining 
the count rate, they showed that the relation from equation \ref{basiccoi} also applied 
to the flat illumination.
We therefore explored a coincidence loss area for correcting the grism spectra.
    
In spectra the detector illumination extends over many pixels, far more than the three 
used in the centroiding of events. This will tend to increase the coincidence loss. 
The physical reasoning behind this is that the centroiding of a photon splash in a 
frame will pick the highest peak. Statistical noise, and the linear spectrum 
in the grism images implies that in some frames neighbouring pixels will have more 
chance to cause coincidence loss than would happen in a point source. Based on this 
simple picture we would expect a larger coincidence loss area for grism spectra than 
for a point source. 

Some new questions arise that were of no concern for a point source or even background 
illumination. One such question is how 
far the effects of coincidence loss reach along and across the spectrum? 
How should we interpret the extent along the spectrum over which the coincidence 
loss works in a bright emission line? 
That distance may be an indication of the extent of the coincidence loss area along the spectrum.
 
\begin{figure}
\includegraphics[width=88.0mm,angle=0]{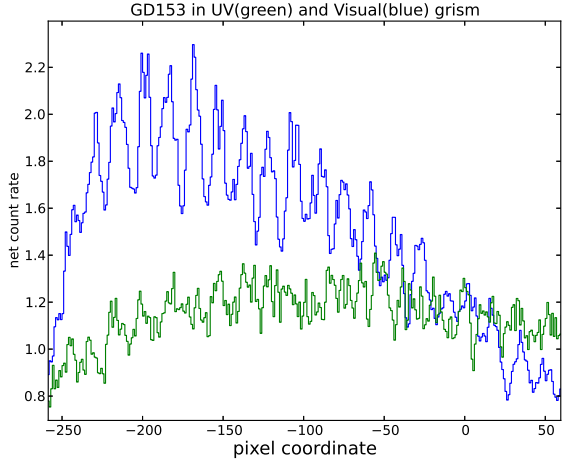}
\caption{The effect of coincidence loss on a smooth input spectrum. The blue 
spectrum shows GD153 in the \visG, the green spectrum in the \uvG. 
}
\label{fig_coi_wiggles}
\end{figure}

When we consider the spectra of a series of sources with increasing brightness, it 
becomes clear that spectra exceeding a certain brightness level show instrumental 
peaks and valleys. 
It is thought that the high brightness affects the on-board centroiding which has 
been fine tuned for a smoother brightness variation. 
The photons at high coincidence loss are mis-registered and appear as bright points 
in a \mod8 pattern in the grism image. 
As the spectrum lies at an angle over 
the pixel grid, the brightness distribution of the extracted spectrum of a very bright 
source shows this variation, even when the source spectrum is smooth. 
This effect is most prominent at the bright part of the spectrum where the grism effective 
area peaks. For example, spectra taken from GD153, a DA 
white dwarf star, see Fig.~\ref{fig_coi_wiggles} show this. That spectrum is 
quite typical for the variations in spatial frequency. The pattern is more pronounced 
in the \visg since it is more sensitive than the \uvG.

For the determination of the coincidence loss, we need to average over this pattern
and this pattern determines the minimum length along the spectrum we need to include. 
Bright spectral lines show that the coincidence loss is quite extended. After some 
trials, we decided to use a length $L_{coi}$ along the spectrum determined 
as 24 subpixels divided by the cosine of the typical angle of the 
spectrum on the detector for each of the grism modes. This averages the count rate over 
most of the variations.  

\begin{table}
   \caption{Coincidence loss area box length and width.}
   \label{coilength}

   \begin{tabular}{@{}lcccccl}
   \hline
Grism mode& $L_{coi}$  & $W_{coi}$  & coi-    &  & \\
          & pixels     & pixels & multiplier &  & \\ 
          &            &        &   $m_{coi}$   & \\
   \hline
\ungb   & 27 & 15 & 1.12$\pm$0.05 \\
\ucgb   & 29 & 15 & 1.11$\pm$0.05 \\
\visnb  & 28 & 14 & 1.13$\pm$0.03 \\ 
\viscb  & 31 & 13 & 1.09$\pm$0.03 \\
   \hline
   \end{tabular}
\end{table}

The coincidence area $A_{coi}$ is most simply represented by a box of length 
$L_{coi}$ and width $W_{coi}$, multiplied with a correction factor $m_{coi}$, 
named coi-multiplier, 
\begin{equation}
\label{coi_area}
A_{coi} = L_{coi} \times W_{coi} \times m_{coi}. 
\end{equation}
The measured count 
rate for the coincidence loss correction is thus determined for that area, and the 
corresponding background is computed for an equivalently sized area. 

Equation \ref {basiccoi} defines a relation between the observed count rate and 
the actual photons incident on the detector.  
We used sources with a known spectrum to fit the data to the theoretical relation in order 
to determine the best coi area width and the corresponding coi-multiplier. 
Data from multiple sources were used to increase the range of brightnesses. 

For studying the coi-effect the spectra were split into 
adjacent areas extending over a length $L_{coi}$. A range of width $W_{coi}$ were chosen.
For each area the average observed count rate was determined. 
That count rate differs from the count rate determined within the 
aperture of the spectrum, as it is used for calculating a correction factor for 
the coincidence loss only. We refer to this as the ``observed coi count rate".  

\begin{figure}
\includegraphics[width=88.0mm,angle=0]{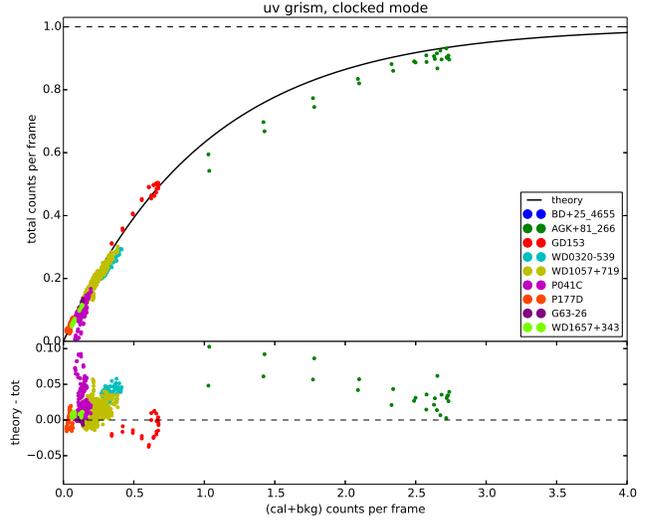}
\caption{The coincidence loss in the \uv grism, clocked mode. 
Using our calibration sources we can find the count rate per frame expected 
to be incident to the detector and plot the observed count rate per frame as  
function of the expected rate. The observed rate is the sum of source and background 
and so we also add the observed background to the predicted rate to be consistent. This plot 
shows the best fit of the observations to the relation of equation \ref{basiccoi} 
using
the parameters in Table \ref{coilength} 
}
\label{fig_coiobs_160}
\end{figure}
\begin{figure}
\includegraphics[width=88.0mm,angle=0]{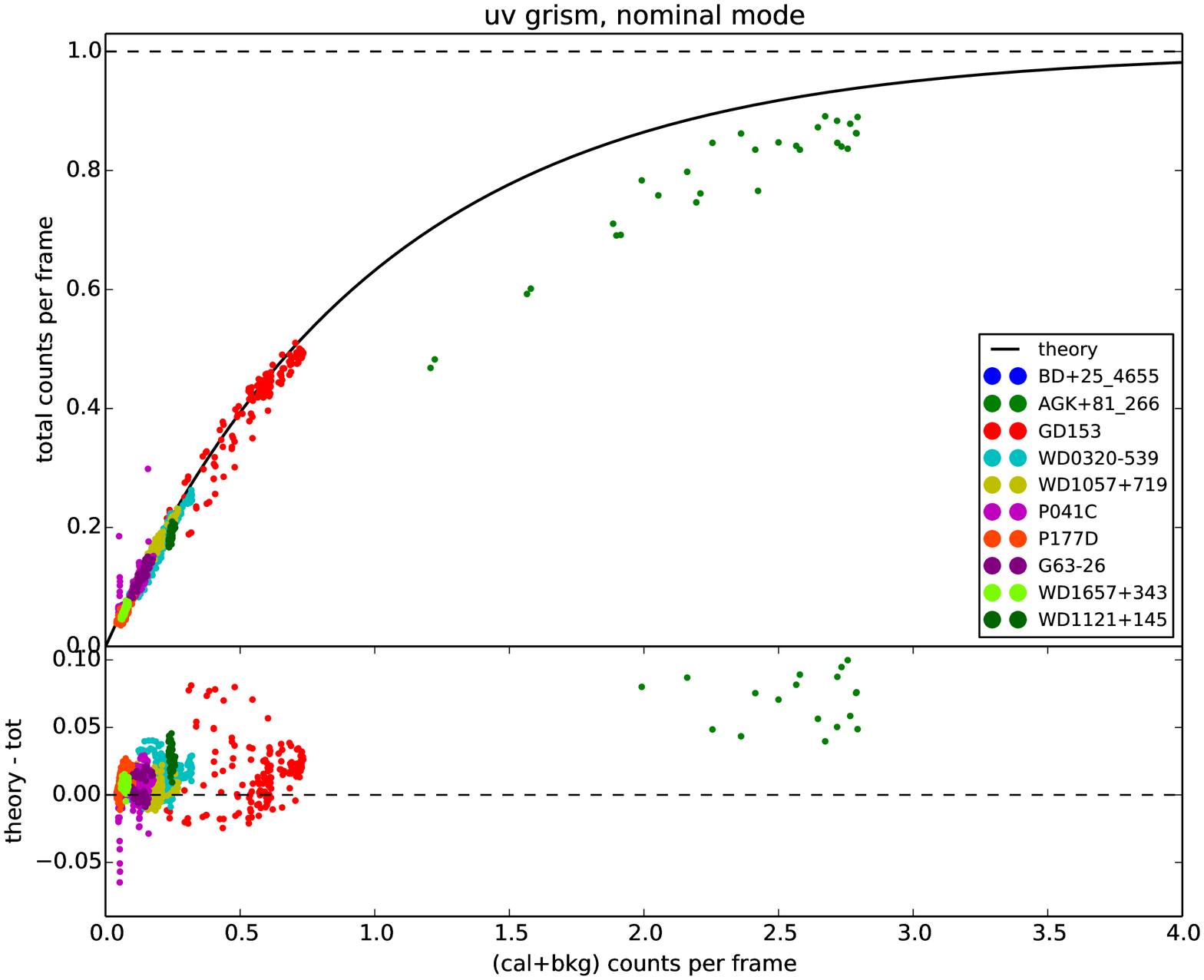}
\caption{The coincidence loss in the \uv grism, nominal mode.
See the caption of Fig.~\ref{fig_coiobs_160}. 
}
\label{fig_coiobs_200}
\end{figure}
\begin{figure}
\includegraphics[width=88.0mm,angle=0]{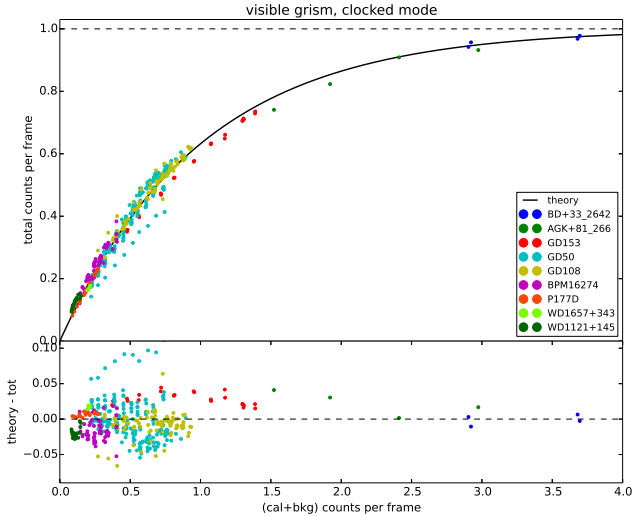}
\caption{The coincidence loss in the visible grism, clocked mode.
See the caption of Fig.~\ref{fig_coiobs_160}. 
}
\label{fig_coiobs_955}
\end{figure}
\begin{figure}
\includegraphics[width=88.0mm,angle=0]{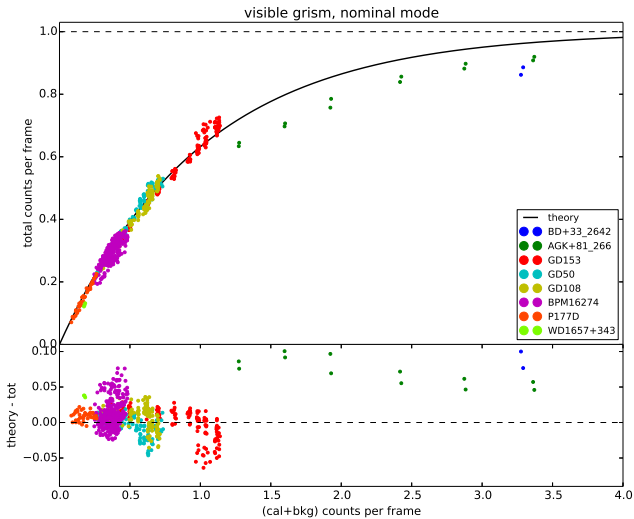}
\caption{The coincidence loss in the visible grism, nominal mode.
See the caption of Fig.~\ref{fig_coiobs_160}. 
}
\label{fig_coiobs_1000}
\end{figure}

The sources that were used have been listed in Table \ref {target_table}. 
Bad areas of the spectra were filtered out, and areas of possible second order overlap 
were removed. All spectra were in the central area of the detector ($800\times800$ 
pixels). 

The reference calibration spectrum of each source is then used as follows to find 
a predicted count rates: i) Using the observed spectrum, wavelength ranges
for each coi-area are determined. ii) The spectral dispersion is used to convert the 
flux per angstrom in the reference spectrum to the flux per channel in the coi-area. 
iii) The effective area is used to convert to count rate per channel. 

Theory tells us that the maximum observed
rate per frame approaches one asymptotically\footnote{Very bright sources with 
incident count rates per frame larger than 6 suffer additional losses when the 
detector PHD saturates in neighbouring pixels.}. 
For that reason we observed some sources that 
were bright enough to reach the limit. The grism images of these 
sources exhibit the \mod8 pattern in the spectrum, and although they need to be 
``averaged" over the coi-box, 
are good enough to determine the coincidence loss. In point sources, incident 
count rates of 3.5 counts per frame leads to an observed rate per frame of 0.97. This is 
considered to be the practical limit for a good measurement. 
For the grisms we see  
similar numbers, but for this analysis incident count rates per frame of 
up to 5.5 were included. 
For each coi width $W_{coi}$ the optimal value for the coi-multiplier 
was determined using
a least squares fit on the individual data points and the $\chi^2$ of the fit was 
computed. We also grouped the data per source, and calculated the coi-multiplier for each 
source by inverting equation \ref{coi_area}. Outliers were clipped in both calculations, 
and data at the ends of the spectra, which have less sensitivity, 
were also not used. Both methods prefer small 
widths for $W_{coi}$ and similar coi-multiplier $m_{coi}$ for each grism mode, see 
table \ref{coilength}. The error on $m_{coi}$ found from the individual
least squares fits was an order of magnitude smaller than the error on $m_{coi}$ derived 
after determining a coi-multiplier per source. We attribute that to systematic errors 
in the measured count rates that depend on things such as how  crowded the field of 
the calibration source is.

For the best fitting values of $W_{coi}$ and $m_{coi}$ we plot in Figs. 
\ref{fig_coiobs_160}, \ref{fig_coiobs_200}, \ref{fig_coiobs_955}, and \ref{fig_coiobs_1000} 
the total observed count rate per frame as function of the predicted count rate 
per frame, where the predicted count rate per frame 
from the source includes the observed background count rate per frame corrected for 
coincidence loss using equation \ref{basiccoi} 
using the same coi-area. The observed count rates level off to a rate of 1 count per 
frame when the incident counts per frame reach around 4 to 5.  
The values found 
for {\em AGK+81 266} tend to be offset from the other data, like {\em BD+33 2642}
which may be due to nearby spectra of bright field stars in their images.  

For continuum sources the coincidence loss affects the shape of the spectrum once the 
incident counts per frame exceed 1, i.e. photons are not measured at the right wavelength 
due to the centroiding error at high illumination. 

To apply the coincidence loss correction to the flux, 
we determine for each channel by what factor, called the coi-factor $f_{coi}$, 
the measured net count rate needs to be multiplied as follows:
\begin{equation}
\label{coi_factor}
f_{coi} = {{\varphi}_{in} \over {\varphi_{m}}}. 
\end{equation}
For determining $f_{coi}$, $\varphi_{m}$ is found by measuring the 
count rate in the coi-area multiplied by $m_{coi}$ (see Eq. \ref{coi_area}) 
and ${\varphi}_{in}$ is then derived using Eq. \ref{basiccoi}. 
For each spectrum channel this factor is computed. 

Once the count rate in a channel of the spectrum is measured using a 
width expressed in terms of $\sigma$ (Eq. \ref{sigma_eq}) as described in 
Section \ref{x-profile}, it is 
multiplied by the coi-length, the coi-multiplier, and 
an appropriate aperture correction. 
Finally we find the incident, coi-corrected count rate for 
that channel by multiplying the result with $f_{coi}$. 

\begin{figure}
\includegraphics[width=88.0mm,angle=0]{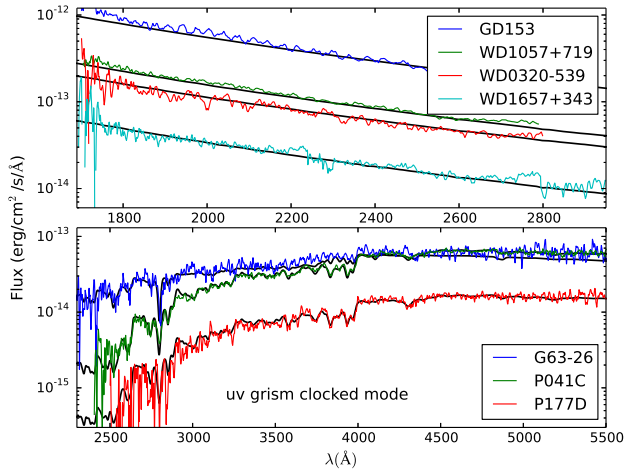}
\caption{Observed, coincidence loss corrected, spectra  in the \ucG. 
The black lines are for the reference spectra. 
The top panel shows three white dwarfs of a range of brightness, the 
bottom panel one F0V subdwarf, and two main sequence solar-type stars. 
The spectra are the weighted average of spectra from 2-7 exposures.
Coincidence loss varies from 4\% to about 70\%. }
\label{coi-spectra-160}
\end{figure}

\begin{figure}
\includegraphics[width=88.0mm,angle=0]{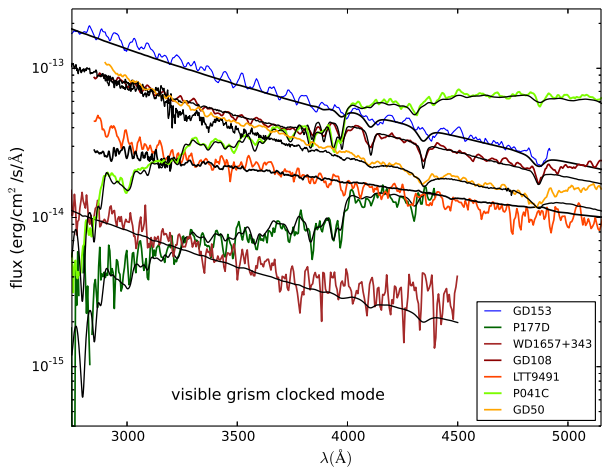}
\caption{Observed, coincidence loss corrected, spectra  in the \visC. 
The black lines are for the reference spectra. 
The P177D and WD1657+343 spectra are from single exposures; the other spectra are 
weighted averages from multiple exposures.
The second order affects the spectra above $\lambda=5200$\AA, and is more pronounced 
for the brighter spectra. The GD153 spectrum is distorted by spatial \mod8 effects of 
the coincidence loss and illustrates the effective brightness limit. 
GD50 which is slightly less bright is also affected to some extent. 
 }
\label{coi-spectra-955}
\end{figure}

Fig. \ref{coi-spectra-160} shows the \uv spectra after the 
coi correction has been applied, where they are compared to the reference 
spectra. 
In Fig. \ref{coi-spectra-955} a similar comparison is made using the 
calibration spectra in the visible grism.

\section{Determination of the Effective area}

\label{effective_area_default}

The conversion from measured count rates to flux 
is done by expressing the detector sensitivity as 
an {\em effective area} which depends 
on the wavelength. The  effective area ${A_{eff}}$ is defined as follows:
\begin{equation}\label{ea_eq}
{A_{eff}} = {hc \over \lambda} \times {{C_{r}^{in} \times r(t)}\over F_{cal} \times (\Delta{\lambda}/bin) }
\end{equation}
Here $h$ is Plancks constant, $c$ the velocity of light, $\lambda$ the wavelength. 
The ($\Delta{\lambda}/bin$) factor is derived from the dispersion relation. 
The known flux $F_{cal}$ 
is obtained from the source listed in Table \ref{target_table}.  
$C_r^{in}$ is the observed source spectrum count rate per spectral bin after 
correcting for coincidence loss and subtracting the coi-corrected background by using 
Eq. \ref{coi_factor}, i.e. the true count rate incident on the detector.  
$r(t)$ corrects for the sensitivity loss of the detector over time, taken to be 
1\%/yr.

\subsection{The effective area for the centre of the detector}

\begin{figure*}
\includegraphics[width=132.0mm,angle=0]{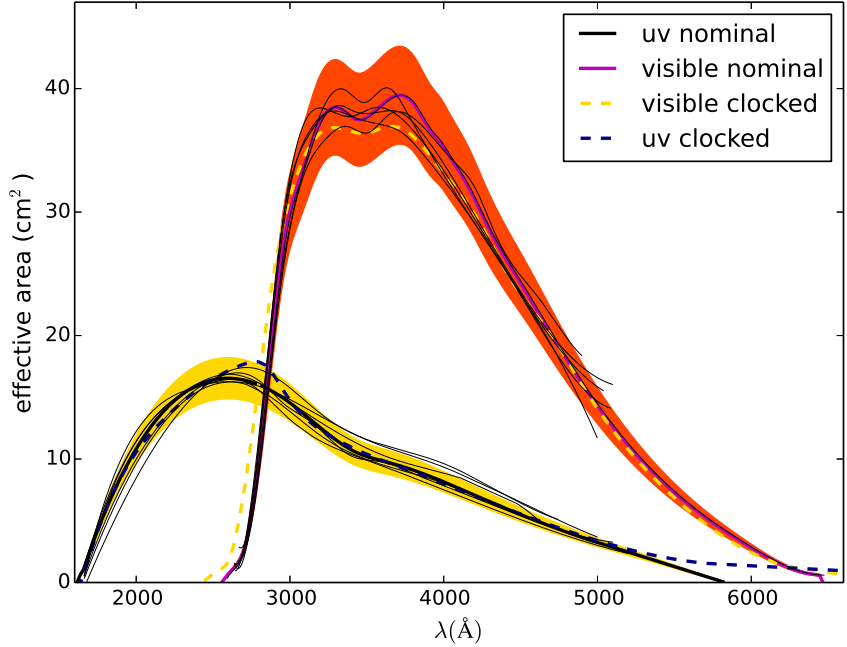}
\caption{The effective areas for the nominal mode. 
Shaded areas show the 1$\sigma$ errors. 
Thin lines show the samples of the effective area from 
different locations on the detector. The dashed curves 
show the clocked grism effective areas at the default 
position for comparison.}
\label{nominal-coi-eff-areas}
\end{figure*}

In order to determine the effective area, the calibration sources in Table \ref{target_table}
were observed several times for each grism mode, but in various detector locations.  
The count rate spectra were extracted with a 2.5~$\sigma$ spectral track width 
(see discussion in Section \ref{aperture_correction}), following the  curvature of the track. 
Where spectral order overlap would be an issue, the data were discarded. 
Areas affected by underlying zeroth orders of field stars were also discarded, 
as well as spectra which had another first order of a different field star overlapping.  
For a good effective area, we need both a blue star and a red star spectrum.
The second order in the red spectrum sets in at a much longer wavelength and therefore allows 
the determination of the effective area to longer wavelengths. 
In practice, in the \uvg the white dwarf spectra were thus used typically for 
wavelengths of $\lambda1650-2900$~\AA, and 
spectra from the $F0$ stars for $\lambda2900-5000$~\AA\  and in the 
\visg the WD spectra up to 4900~\AA\  and the $F0$ stars beyond. 
In the \uvg\ the spectra of the $F0$ stars suffer contamination from order 
overlap for $\lambda > 5000$~\AA.  
The correction for coincidence loss was applied to all spectra. 

Initially, the effective areas were derived for the faintest sources at the default 
(centre)  
positions primarily to minimize the effects of coincidence loss. 
The faintest sources, WD1657+343, and WD1121+145, have count rates less 
than or comparable to the background. GSPC P177-D, GSPC P 41-C, WD0320-539,  
are only slightly brighter, so their coi-correction is still considered 
small  in the \uvg but is unfortunately larger  in the \visG, see Table 
\ref{weak_ea_cal}.

\begin{figure}
\includegraphics[width=88.0mm,angle=0]{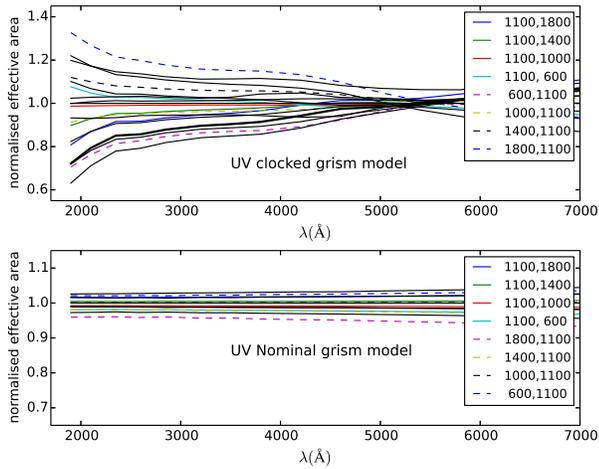}
\caption{The variation of the normalised effective area as predicted by the 
model for spectra with anchors at various locations on the detector. 
Effective areas are plotted for the set of anchor positions with X=1100,Y varying 
and Y=1100, X varying, to illustrate the variation for the clocked (upper) 
and nominal (lower) grisms.
}
\label{normalised_model_flux}
\end{figure}

Many calibration spectra used here were taken early in the mission, especially 
for the \visG, and so do not have a corresponding image in one 
of the lenticular filters taken right before or after the spectrum. 
For those, if possible, the accuracy of the wavelength scale was
corrected prior to the derivation of the effective area using the 
spectral features. 
Without doing this correction, dividing by the reference spectrum $F_{cal}$  caused 
large excursions in the effective area near spectral lines. 

Finally, the effective areas derived from individual spectra and from each source 
were summed,  weighted by the error on the data.  The resulting effective areas
were fit by a smoothing spline to remove noise. 
The brighter sources were included next, and it was confirmed that they gave 
an effective area consistent with that of the weak sources alone, but with a 
smaller effective area error.
The resulting effective areas are shown in Fig. \ref{nominal-coi-eff-areas}.

The accuracy varies by location on the detector and by detector mode, 
and the available total exposure in the calibration 
spectra, see Table \ref{ea_error_table}.
The smaller effective area at shorter and longer wavelengths
means the errors in $A_{eff}$ grow larger as a percentage of the values. 

\begin{table}
   \caption{Flux calibration sources with low coincidence loss used 
   for the initial effective area determination at the default position. 
   See also Table \ref{target_table}.}
   \label{weak_ea_cal}

   \begin{tabular}{@{}llcrl}
   \hline
Grism mode& source name& number of&coincidence \\
 
& & spectra&loss \\
   \hline
\ungb&WD1657+343 &8 & $\le 5\%$ \\
\ungb&WD0320-539 &8 & $\le 9\%$ \\ 
\ungb&WD1057+719 &7 & $\le 9\%$ \\ 
\ungb&GSPC P177-D &14 & $\le 5\%$ \\
\ungb&G63-26 &3 & $\le 10\%$ \\ 
\hline
\ucgb&WD1657+343 &4 & $\le 5\%$ \\ 
\ucgb&WD0320-539 &7 & $\le 9\%$ \\ 
\ucgb&WD1057+719 &8 & $\le 9\%$ \\ 
\ucgb&GSPC P177-D &2 & $\le 5\%$ \\
\ucgb&GSPC P 41-C &2 & $\le 7\%$ \\
\ucgb&G63-26 &2 & $\le 10\%$ \\
\hline
\visnb&WD1657+343 &2 & $\le 10\%$ \\ 
\visnb&GSPC P177-D &1 & $\le 10\%$ \\
\hline
\viscb&WD1657+343 &1 & $\le 13\%$ \\ 
\viscb&GSPC P177-D &1 & $\le 14\%$ \\

   \hline
   \end{tabular}
\end{table}

\begin{figure*}
\includegraphics[width=172.0mm,angle=0]{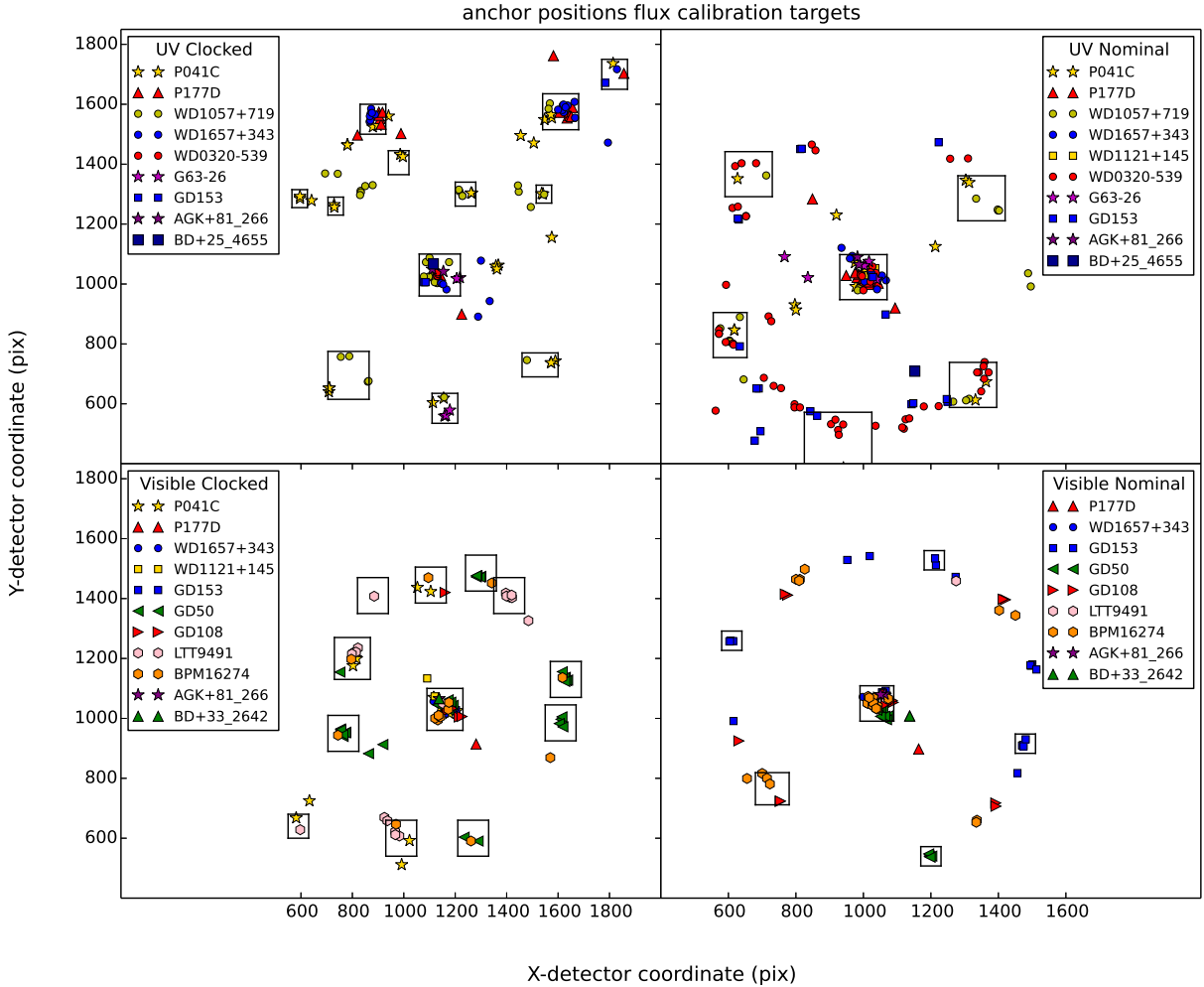}
\caption{The location of the anchors for all the flux calibration 
spectra. The boxes indicate spectra which have been averaged together
for that position.  }
\label{offset_obs}
\end{figure*}

Note that at the default positions the effective areas for the nominal 
and clocked grism modes for each grism are nearly identical, 
as would be expected since they are formed by the same optical elements.

\subsection{The effective area over the whole detector}

It is useful to first consider the model predictions for the variation 
of the throughput over the detector. 
For a spectrum at an offset position the model can be used to predict
the ratio of the effective area to that on-axis. 

For the nominal mode of each grism, the throughput is predicted to vary slowly, 
typically by less than 5\% from the on-axis value over the detector. 
However, for the clocked grism modes the throughput varies much more. 
Fig. \ref {normalised_model_flux} shows the results for the \uvG. 
Positions were varied for fixed X and fixed Y anchor position to 
illustrate this variation. 

\begin{figure*}
\includegraphics[width=172.0mm,angle=0]{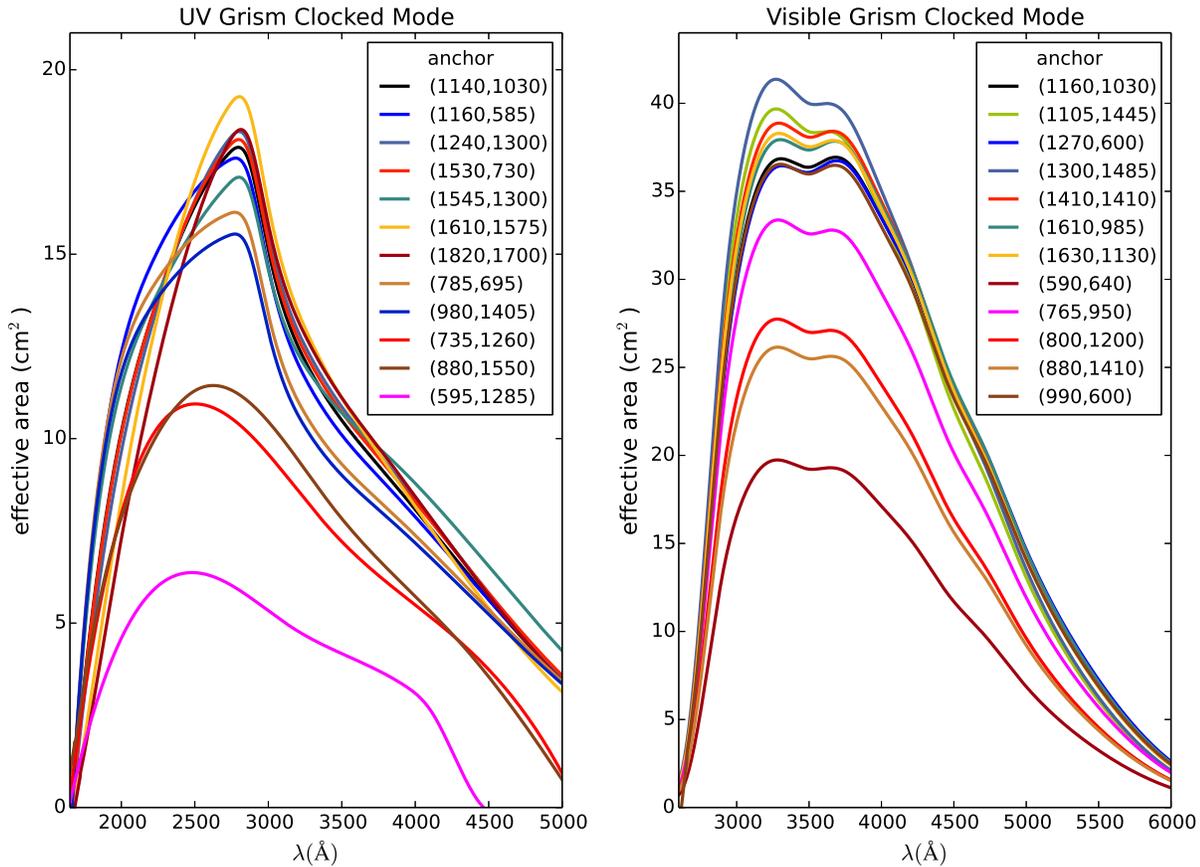}
\caption{The effective areas for the clocked modes at various offsets. 
The offset is labeled according to the anchor position in detector coordinates. }
\label{offset_ea}
\end{figure*}

For the effective area determination over the whole detector  
all sources, including  those with a larger coincidence loss were used, 
after making the coincidence loss correction described in Section \ref{coi2}. 

The locations on the detector where calibration spectra were taken are shown in 
Fig. \ref{offset_obs}, with one panel for each grism mode. Notice that the 
\ucg includes a fair number of observations in the upper right hand corner, 
where the second order is shifted away from the first order, allowing the 
use of the first order up to 4500\AA.  
Additional observations were made in the upper left corner, where the 
sensitivity is found to drop off in the clocked modes.
Also, there are proportionally more spectra from a larger variety of 
calibration sources taken near the default position. 
For each grism mode, about 100 spectra from a range of sources were available 
for the calibration. 

The positioning of the early calibration spectra did not 
use a "slew-in-place"\footnote{A slew in place means the spacecraft is commanded to do a 
second positioning refinement which enables the accurate positioning of the spectrum on the 
detector to within about 20 pixels}, and therefore were not always found at the 
intended positions.  

Spectra that are near to each other and contain both blue and 
red spectra can be grouped to give an effective area.
We find that the variation of the effective area 
as measured at different positions for the nominal grism modes 
is consistent with the model predictions within the uncertainties.  
For that reason we have decided to treat the effective areas for the 
nominal grism modes to be constant over the detector within the quoted error. 

Based on differences between blue (WD) and redder (F0)
calibration sources, possible contamination due to the second order 
contribution increase from  20\%  up to
80\% between 5000 and 6600~\AA\ in the \visN.  

For the clocked grism modes, the variation of the effective area 
over the detector is predicted to be smoothly varying. 
However the observed effective 
areas in some locations were noisy due to limited data.
To remove this noise, the ratio of the effective area to that at the default position 
was constrained to be a smooth polynomial fit. 
The resulting effective area varies smoothly over the detector. 

In Fig. \ref{offset_ea}, for the clocked modes, 
effective areas are shown for locations 
where sufficient observations were available.  
The position of the effective area on the 
detector is indicated by the anchor position given in the legend.
At a glance one can see that in the clocked grism modes the 
effective areas for anchor positions with small  X-coordinate are
markedly smaller, and have a cut-off at some wavelength toward the 
red.
The variation in sensitivity and the cutoff at long wavelengths for 
the left hand side of the detector was qualitatively 
confirmed by the model for the clocked grism. 

It is useful to make a comparison to the effective area of the \uvot {\it white  filter} 
(see Fig. \ref{fig_effarea_lam}) 
which is transparent. The \visg effective areas for the wavelengths above 3000~\AA\ 
are smaller and fall off just like in the {\it white filter} \citep{poole}; 
also the bumps at the peak correspond in wavelength 
to those seen between 3000 and 4000~\AA\  in the {\it white filter}.

\begin{table}
   \caption{Typical errors in the effective area.}
   \label{ea_error_table}

   \begin{tabular}{@{}llccrl}
   \hline
UV   & anchor&  \% error at & \% error & \% error & notes \\
Grism&       & $1750\AA$    & (1,3) &(2,3) \\    
   \hline
nominal&[1000,1080]  & 30 &15 &15 &(4) \\
\hline
clocked&[1140,1030]  & 19 & 9 & 9 &(5) \\ 
clocked&[1160, 585]  & 12 & 6 & 9 \\ 
clocked&[1240,1300]  &  9 & 6 & 9 \\ 
clocked&[1530, 730]  & 11 & 6 & 9 \\ 
clocked&[1545,1300]  &  9 & 7 & 9 \\ 
clocked&[1610,1575]  & 18 & 9 & 9 \\ 
clocked&[1820,1700]  & 28 & 5 & 9 \\ 
clocked&[ 595,1285]  & 20 & 9 & 9 \\ 
clocked&[ 735,1260]  & 20 & 9 & 9 \\ 
clocked&[ 785, 695]  &  6 & 7 & 9 \\ 
clocked&[ 880,1550]  & 13 & 7 & 9 \\ 
clocked&[ 980,1405]  & 20 & 9 & 9 \\ 
\hline
Visible& anchor& \% error at&\% error&\% err &notes \\
Grism  &       & $6000\AA$  & (1,3)  &(2,3) \\    
   \hline
nominal&[1050,1100] & 13 & 11 & 11 &(4) \\ 
\hline
clocked&[1141,1030] & 11 & 15 & 15 &(5)\\
clocked&[1630,1130] & 11 &  9 & 15 \\
clocked&[1270, 600] & 11 &  9 & 15 \\
clocked&[1300,1485] & 11 &  7 & 15 \\
clocked&[1410,1410] & 11 & 10 & 15 \\
clocked&[1105,1445] & 11 & 10 & 15 \\
clocked&[1610, 985] & 11 &  9 & 15\\
clocked&[ 765, 960] & 11 &  6 & 15 \\
clocked&[ 880,1410] & 11 & 10 & 15 \\
clocked&[ 800,1200] & 11 & 15 & 15 \\
clocked&[ 990, 600] & 14 & 6  & 15 \\
clocked&[ 590, 640] & 30 & 13 & 15 \\
   \hline
   \end{tabular}
   \begin{tabular}{@{}lll}
   (1) & RMS error from consolidated EA for each source. \\
   (2) & recommended error. \\
   (3) & errors based on where $A_{eff}$ is larger than half the \\
       & maximal value. They will be larger at wavelengths with \\
       & lower effective area. Typically that applies to the \\
       & ranges 1950-4400~\AA(\uv) 2850-4800~\AA(visible)\\
   (4) & effective area at $\pm$600 pixels from the default position .\\
   (5) & effective area at $\pm$80 pixels default position.\\  
   
   \end{tabular}
\end{table}

\subsection{The accuracy of the effective area calibration}

In Table \ref{ea_error_table} the errors from the determination of the effective area
have been listed. 
The error due to the accuracy of the calibration spectrum, and the 
Poissonian noise in the spectrum and background is typically 3-4\% after averaging over 
about 5-10 spectra, and has not been listed for each effective area result. 
The combined RMS error derived for each wavelength from the effective area from all 
relevant spectra combined is listed in the third and fourth columns. 
The third column lists the value at a fixed wavelength outside the mid-sensitivity 
range, while the 
fourth column lists the maximum value at the middle of the sensitivity range.
For the \uvg the sensitivity is already quite low at 1750~\AA\  and this wavelength has been
given specifically to give a measure of the largest error expected in the blue 
part of the \uv range where the sensitivity gets low. 
For the visible grism, the sensitivity at the short wavelengths drops off 
so rapidly, that the error given for the range in column 5 is often that at 
around 2850~\AA\ where the effective area is about half the peak value. The value 
at 6000~\AA\  has been chosen since the spectrum at longer wavelengths 
has a low effective area. 
For all effective areas the errors reach a minimum value of around 5-8\% near the 
peak effective area. 

To understand the effect of the error introduced by coincidence loss in our 
effective area determination, we considered the following.
The error in the coincidence loss is proportional to the error in the count rates 
shown in Figs.  \ref{fig_coiobs_160}, \ref{fig_coiobs_200},
\ref{fig_coiobs_955}, \ref{fig_coiobs_1000}, so typically a 5-10\% error 
is introduced. 
The magnitude of the coincidence loss correction depends on the calibration 
source used, and so does its error. 
For sources less bright than the background, the accuracy of the background determination
is a source of error. The background changes when the roll angle of an observation 
varies due to unresolved zeroth orders and is difficult to measure. 
However, the error is unlikely to exceed 10\% 
or the observed variation in count rates for weak sources would be larger. 
The other factors contributing to the error in the effective area are the 
random measurement errors and the uncertainties in the flux standards.
At some detector locations only a very limited number of calibration observations
were available to derive the effective area and variance between the observations  
is found to be lower than in areas with many spectra from different sources. 
In that case there is no proper sampling of the variance in the measurements.

Finally, considering all these effects, in the fifth column the 
recommended uncertainty in the effective area is listed.    
For the nominal grism effective areas which were determined at 
various offsets, the accuracy was such that these were deemed 
consistent with the effective area at the default position. 

The effective area curves in the \uvg above 4900\AA\  may have 
been affected by some second and third order emission, which 
tends to increase the resulting effective area. 
This contaminating emission is present above 5000\AA\  in 
the F0V spectra that were used.  
This would be especially so for the effective areas with order 
overlap, but not for those in the \ucg at the large offset at 
high anchor (X,Y) coordinates. 
The effective areas all show a similar slope, so that the effect 
is thought to be less than 20\% above 5000\AA. 
In the \visg the second order is much weaker since the grism 
was blazed. The only indication of second order contamination  
is that the flux in WDs shows some enhancement 
above 4900~\AA. For the effective area above 4900~\AA\  only P177D and
P041C of spectral type F0 were used which are not expected to 
have second order contamination in the \visG.  

\subsection{Effective area for the second order}
\label{EA2nd}

There are two ways in which the second order count rate can be measured 
and separated from the first order count rate. 

The first method is where orders completely overlap, 
and given an accurate effective area for the first order, the second order 
contribution can in principle be found by subtraction. 
The correction for coincidence loss treats the total flux of first and 
second order combined, and this has to be used appropriately in the analysis.
The second method is at a large offset position in the \ucG. 
There the second order separation from the first order  
allows us to measure the count rate of that separated part of the second order. 
Here also a good wavelength calibration for the second order is needed. 
The noise in the measurement requires a reasonably bright source.
However, the source cannot be too bright since then the first order will affect the 
second order by coincidence loss. 
We will only consider the first method here.  

\begin{figure}
\includegraphics[width=88.0mm,angle=0]{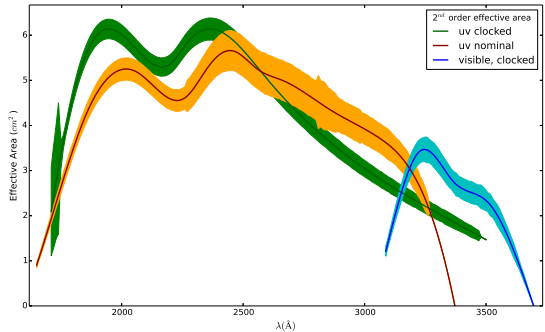}
\caption{The effective areas of the second order in the \uvg and \visg at the 
default position. The effective area is shown as function of the second 
order wavelength. Shaded regions indicated the uncertainty in the effective area
of the second order.}
\label{2nd_order_EAs}
\end{figure}

To derive the second order effective area at the default position we used the observed 
rates after subtracting the background, and applying the coincidence loss 
correction. The rate due to the first order was then derived using the 
reference spectrum and the known first order effective area. This rate 
was subtracted to obtain the second order rates, already corrected for 
coincidence loss. The second order rates were subsequently used in equation
\ref{ea_eq} to derive the second order effective area, limited to the 
default cross-dispersion width of the first order. This 
is the most useful measure of the size of the second order contamination 
affecting the first order.    

We examined the results for the various flux calibration sources, and found that 
GD153 and WD1057+719 gave results which were consistent with the much noisier 
data from the other sources. 

The error in the second order is rather large, as it includes a subtraction 
of the first order effective area, and the noise is also larger for that reason.
For the first order effective area error a conservative 11\% was adopted. The 
error on the second order effective area for an individual spectrum then ranges 
from 20 to 40\%.  The derived effective areas were averaged over multiple spectra which 
reduces the error to that shown in Fig. \ref {2nd_order_EAs}.   

In the \ucg the second order becomes displaced normal to the dispersion direction with 
the size of displacement dependent on position of the spectrum on the detector. 
In these cases a different part of the second order will contribute to the first order. 
Estimating the effective area of the second order to compute the contamination
in the first order is thus only providing an approximate upper limit for spectra away from the 
default centre position. In the \visg the orders overlap over the whole detector and this
issue does not occur. 

Comparison to the first order effective area in Fig. \ref{nominal-coi-eff-areas} 
shows that the second order effective 
area in the \uvg is less than half that in the first order for equivalent 
wavelength, 
but in the \visg the second order effective area is less than 10\% of that in the 
first order which means that usually it can be disregarded except where the first 
order effective area is very small, i.e., at long wavelengths. 
In the \uvg the first order effective area at long wavelengths is overtaken by 
the second order effective area, resulting in equivalent effective areas
on the detector at around 4800~\AA\ in first order and 
at 2600~\AA\  in the second order, after which the second 
order effective area is larger than that of the first order.

\section{The zeroth order  in the \uv grism nominal mode}


\subsection{Zeroth order effective area}
\label{zo_ea}

The zeroth order dispersion as predicted by the model can be fit approximately 
by a rational function as:
\begin{equation}\label{zerodisp}
  \lambda = 1220. - 110715/(p-35.6)\mskip+12mu  {\rm \AA}, 
\end{equation}
where $p$ is the pixel distance to the anchor point. 
The anchor point of the zeroth order is not directly measurable, but is 
selected to be close to the peak of the zeroth order emission.

As we show later, the zeroth order in the \uv grism provides a similar 
response to the $b$ band filter.

The dispersion relation can be used to predict the wavelengths that fall on a pixel. 
When doing that, it is clear that at the longer wavelengths a large part of the spectrum 
falls on only a few pixels, while the \uv part of the spectrum is 
spread out into a relatively large tail.  
Of course, the peak emission will be spread out over several pixels, since the 
instrumental PSF is about 5 pixels FWHM for the zeroth order. 

The calibration spectrum was compared to the measured zeroth order counts. Since 
the red part of the spectrum falls on such a few pixels, a faint white dwarf spectrum is ideal 
for calibrating the zeroth order since its flux rises in the blue and is 
faint, so that coincidence loss is very minor. 

The natural way to proceed 
is to convert the CALSPEC WD 1657+343 spectrum  $F_{cal}(\lambda)$ as a 
function of wavelength to a count rate spectrum as a function of 
pixel distance to the anchor point, which we define as $C_{cal}(p)$, 
with the anchor point chosen appropriately. 
\begin{equation}\label{xxx}
   C_{cal}(p) = \mskip-6mu\int\limits_{{\lambda}1}^{{\lambda}2}\mskip+18mu 
   F(\lambda) (\lambda/hc) \rmn{d}{\lambda},
\end{equation}
where ${\lambda}1$ and ${\lambda}2$ are bounding wavelengths of the pixel $p$.
The spectrum is then converted to count rate by dividing the flux by the photon energy. 

\begin{figure}
\includegraphics[width=88.0mm,angle=0]{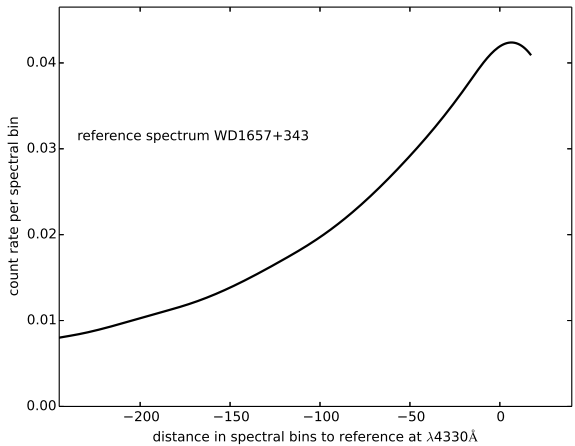}
\caption{The white dwarf reference spectrum of WD1657+343 
converted to counts per spectral bin per $cm^2$ for the 
adopted zeroth order dispersion. The size of the spectral bins increases 
hyperbolically with wavelength. This shifts the peak to the red. }
\label{fig_calspec}
\end{figure}

In Fig. \ref{fig_calspec} the calibration reference spectrum of WD1657+343 is shown
as it should appear in zeroth order. 
Although the flux rises to the blue, the predicted counts per bin  
peak in the red near 4350\AA\  due to the nonlinear dispersion. 
The wavelength range inside a single bin increases so much 
towards the red part of the spectrum that it leads to a counter-intuitive WD flux(bin)  
spectrum for those of us who are used to  
the normal flux rising in the blue in a flux(wavelength) plot.  
With the non-linear dispersion of Equation \ref{zerodisp}, 
the anchor at $p$=0 corresponds to $\lambda$ = 4329\AA, which is 
the $b$ band effective wavelength. 
Ten bins up, at $p$ =  +10, the median wavelength of the 
$v$ filter is passed, while at $p$ = +25, the wavelength 
is more than 7000\AA, at which 
point the effective area of the instrument will cut off any further contribution.   

\begin{figure}
\includegraphics[width=88.0mm,angle=0]{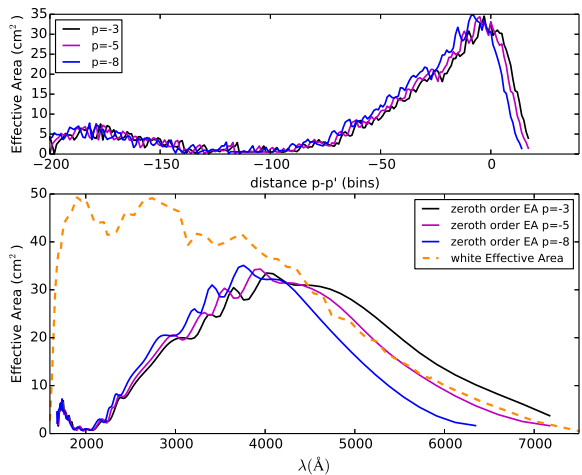}
\caption{The effective area for three values of $p-p'$. The top 
panel shows how the effective area is distributed in the observed spectrum, 
while the bottom panel shows the smoothed effective area as function of wavelength.
Note that in the \uv there is a small response. The effective area of the 
white filter should be larger than that for the grism zeroth order.}
\label{fig_effarea_lam}
\end{figure}

With the effective area definition  based on the counts per pixel using equation \ref{xxx}, 
the expression for the effective area simplifies to 
\begin{equation}
\label{zo_eff_area}
   A_{eff}(p) =  C_{obs}(p)/C_{cal}(p),  
\end{equation}
with $C_{obs}(p)$ the observed count rate per pixel\footnote{Coincidence loss is 
present but neglected here.}, and $C_{cal}$ is the 
count rate per pixel per cm$^2$, see Eq. \ref{xxx}. 
The measurement is really $C_{obs}(p')$ with $p' \ne p$, since we do 
not know where in the spectrum the anchor point really lies, resulting in an 
uncertainty in matching the calibration spectrum 
to the observed spectrum. However, we are aided by the small number of pixels that the red 
part of the spectrum falls on, and also that the effective area drops to zero far short of the 
longer wavelengths covered by the grism if the anchor in the spectrum 
is made to be at too short a wavelength.  
Furthermore, as another check, the white filter, which is 
blank, should have a larger effective area than the grism (see Fig. \ref{fig_effarea_lam}).      

To study the sensitivity of the result for other choices of $p-p'$, several other 
cases were computed.
For our preferred choice the peak in the observed count rate spectrum was 
matched to the adopted anchor point in the calibration spectrum to derive the 
effective area. Then, an offset of several pixels to either side was applied 
and the effective area was rederived. A smoothing spline was applied to remove 
the noise. 
The result shown in Fig. \ref{fig_effarea_lam} gives the effective area 
for three offsets. The reddest peaking 
curve actually has too large an effective area when comparing with that of the 
$white$ filter effective area. Therefore, we exclude the reddest peak from 
further consideration. 
The next offset p=-5 peaks at 3920\AA, and closely matches on the red side the effective area 
of the $white$ filter. The p=-8 is perhaps the best match with the effective area below that 
of the $white$ filter, with a peak at 3740\AA\  
while also the shape conforms better to the $white$ filter.  
The \uvot $u$ filter has a central wavelength of 3465\AA, while 
for the $b$ filter it is 4392\AA. 
{\changed With the unknown effects of the 
coincidence loss in the zeroth order its effective area is considered to 
be compatible with either the $u$ or the $b$ band filter. In the next section 
this has been investigated further.}

The \visg has a much weaker zeroth order. Applying the same analysis as for the \uvG, 
we find a peak at $4200 \pm 200$\AA\  with a peak effective area of $6.6\ cm^2$.

\subsection{Photometry with the zeroth order in the \uvg}

The zeroth orders are still present with reasonable signal to noise even
when the first order is too faint to distinguish from the background. 
With the \ung in the automated response sequence typically taking an 
exposure 300-350s into a Gamma Ray Burst, we can still use the 
zeroth orders for the fainter bursts to obtain an equivalent magnitude. 
\label{zo_section}

As we have seen, the zeroth order emission is extended. Sources 
with different spectra and thus different spectral energy distributions 
will tend to peak slightly differently in the zeroth order. However, 
as we have seen in Fig. \ref{fig_calspec}, even from a very blue source 
the redder photons are those most readily detected in the zeroth order.
 
In our method, the adopted approach is to put an aperture which is 
a 10\arcsec (about 20 pixel) radius 
circle at the location of the zeroth order in the grism detector image
using the WCS-S sky position after a successful application of {\tt uvotgraspcorr}.  

For the photometric calibration, the same calibration sources as in \citet{poole} were
used.

\begin{figure}
\includegraphics[width=88.0mm,angle=0]{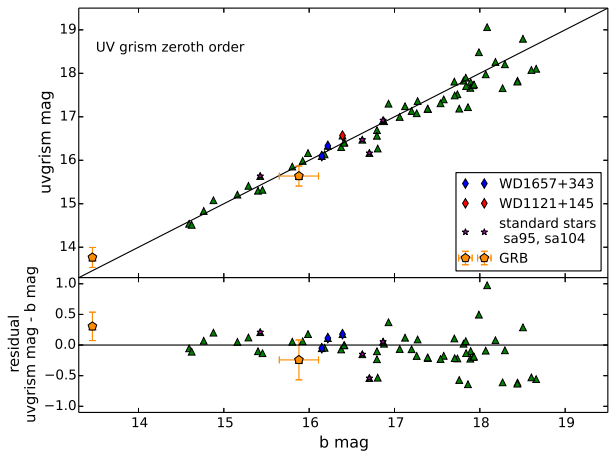} 
\caption{The \uv grism zeropoint magnitude derived using zp=19.46, compared to \uvot $b$.
The GRB data have only been used for verification.}
\label{fig_zp1}
\end{figure}

\begin{figure}
\includegraphics[width=88.0mm,angle=0]{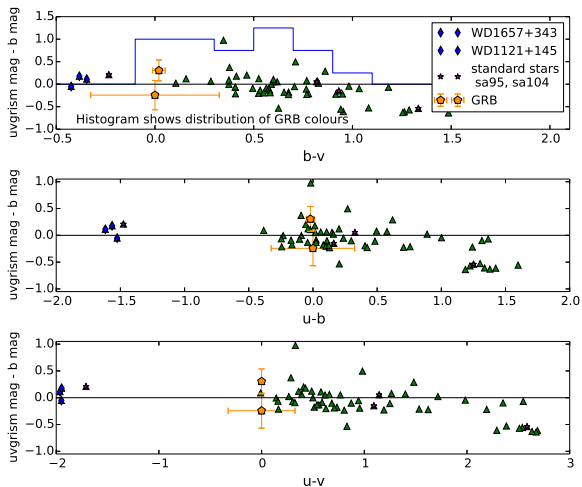}
\caption{The \uv grism zeropoint colour dependence residuals. The GRB data 
have only been used for verification.}
\label{fig_zp4}
\end{figure}

The measured count rate was compared to the photometry in $v$, $b$, and $u$, and showed that 
the photometry is most closely matched to that in $b$, which has an effective 
filter wavelength of 4329~\AA.  Red stars with $u-v < 2$ or $u-b < 1.1$ were excluded.  
The best fit zeropoint is 19.46 for a \uvot $b$ magnitude on the Vega system. 
The two panels in Fig. \ref{fig_zp1} show the \uvg magnitude as determined 
by using this zeropoint versus $b$ magnitudes and the residuals. 
The three panels in Fig. \ref{fig_zp4} show the residuals  
against colour. 

{\changed We found a significant dependence of the grism - $v$ magnitude on colour,  
but not so strong with grism - $b$ magnitudes,   
and for that reason we derive an equivalent
$b$ magnitude from the grism zeroth order. }
For a source which is not redder than $u-v=2$ or $u-b=1.1$ 
then the count rate in a 10\arcsec\  circular aperture in the \uvg  can be used to 
determine the $b$ magnitude  with an error of 0.23 mags RMS.

For verification of the validity for GRBs, the data from GRB081203A \citep{kuin08} 
and GRB081222A were used. The GRB magnitudes and colors were derived from their 
spectra and lightcurves.  The grism magnitude and error were derived as described above 
and they were plotted in Figs. \ref{fig_zp1} and \ref{fig_zp4}. 
It can be seen that the GRB grism magnitudes are consistent with the calibration.


\section{The grism uvotpy software and output file formats}
\label{softw}
This calibration has been incorporated into a software package 
written in the Python language. The basic parts of the software were  
originally written to perform the calibration 
and were converted into a tool for users.
The \uvotpy\  \citep{uvotpy}\  
was released in several steps while the calibration 
was ongoing, and includes at present the latest calibration files. 
It requires the normal \Sw \uvot  software and CALDB which are used 
for example for further processing the data. The code also depends 
on several other Python packages, like 
{\tt matplotlib}\footnote{http://matplotlib.org}.  
The software can be downloaded from the 
UCL/MSSL UVOT website\footnote{see http://www.mssl.ucl.ac.uk/www\_astro/uvot/} .

\label{uvotpy}
The \uvotpy\  writes the extracted spectral data to \fits format files.  
The main file has several extensions. The first extension 
contains the data for use with the 
{\tt XSPEC}\footnote{http://xspec.gsfc.nasa.gov} program,
though the redistribution matrix file is only produced when 
a parameter is set in the call.  
The rates have been corrected for coincidence loss. 
The second extension is of interest to non-XSPEC users and contains in 
a binary table columns, for the first order: the wavelength, 
the net count rate and background count rate uncorrected for coincidence loss, 
the coi-corrected flux, the coi-correction factor that has been applied, 
the channel number, and the aperture correction factor which was applied in the 
calculation of the flux. 
Optionally, the second order wavelength, count rate, and background are supplied, 
and the rates in the coincidence area.
A flag column lists the automatically generated flags. These provide
warnings of, for example, possible nearby zeroth orders. 

Spectral extraction requires at a minimum the sky position in RA and Dec in 
decimal degrees, the observation ID, and the grism detector file extension 
with the spectrum to be processed. Optional parameters control such things as  the 
data directory path, whether the second order should be extracted separately 
(for offset observations where the orders split), and the background 
region(s) for the spectrum.

The program automatically derives the anchor position by using information 
in the (optional) lenticular filter image header(s), 
the grism detector image file header, and calibration files. 
Based on the anchor location the curvature of the spectrum is found, and the 
angle under which the spectrum lies on the detector image. 
A slice around the spectrum is extracted and rotated which involves a
resampling. The rotated image slice is then used to extract the spectrum. 
The count rate within the spectral track is 
extracted. The first and second order are extracted independently, so 
where there is overlap, the counts end up in both first and second orders. 
A plot is made to show the extracted count rates of the first and second orders, 
and the first order count rate spectrum is used to make a prediction of 
the second order count rate.
Comparison of the first and second orders based on pixel position can help 
determine the location of second order features in the first order. 
The program makes a flux calibration by applying the  
coincidence loss correction and effective area, and 
writes the results to the output files. 
A plot shows the flux calibrated spectrum, and a very rough estimate of the 
second order spectrum, if present.

\label{extraction}

Since the width varies slightly from exposure to exposure,
to ensure the same encircled energy in an
extraction, the spectra are extracted after fitting the 
width of the spectral track with a Gaussian curve, interpolating the result 
with a polynomial to remove the effect of \mod8 noise on $\sigma$ and using a 
width of 2.5 $\sigma$.

In more detail, the following steps are used to extract the spectra:
\begin{enumerate}
  \item the anchor position is determined.
  \item the image is rotated around the anchor using the model dispersion 
  angle for the anchor. 
  After the rotation, the zeroth order and \uv part of the spectrum is to the left, the higher orders 
  and longer wavelengths to the right. Due to the curvature the spectrum is generally
  not completely aligned with the x-axis.  
  \item the rotated image is clipped 100 pixels above and below the centre.
  \item the curvature of the spectrum is retrieved.
  \item a correction is made for the distance of the anchor to the centre of the spectral track
  \item a gaussian is fit to the width of the spectral track over 30 pixel slices 
  along the dispersion and 
  interpolated with a polynomial.
  \item the counts are extracted in a curved slit (default halfwidth 2.5 $\sigma$).
  \item the aperture correction is computed.
  \item the flags are populated based on catalogue sources near or on the spectrum.
  \item the slit parameters are saved.
  \item the coincidence loss factor is calculated and used.
  \item the background counts are subtracted.
  \item statistics are calculated.
\end{enumerate}

The background calculation is performed as follows.

\begin{enumerate}
   \item an initial background image is made from the rotated image.
   \item pixels $> 3 \sigma$ above the mean, are replaced  
   with the mean of the $3 \sigma$ clipped image. 
   \item the background image is divided into 80 areas in the dispersion 
   direction (where the background varies in the clocked image).
   \item for each area bright pixels, where (data $> mean+2\sigma$), are replaced 
   with the mean.
   \item the spectrum is smoothed using a boxcar filter 50 pixels along dispersion, 7 pixels 
   across dispersion direction. 
   \item the background is extracted from default regions or as requested.
   \item the background regions are interpolated to match the extraction slit of the spectrum.
\end{enumerate}

The result is a robust background estimate. 
In very crowded fields the background may
include unresolved spectra, while excluding the 
stronger spectra present. 
Zeroth orders are properly masked out as well. 

Flags are maintained for bad or possibly bad areas. 
The  {\tt uvotgetspec.quality\_flags()} defined values are:
`good': 0,
`bad': 1, which means the data is definitely bad;
`zeroth':2, which means there is a zeroth order near the spectrum
`weak zeroth': 4, which means that the zeroth order near the spectrum does not 
have a halo; `first': 8, there is contamination by a first order spectrum; 
`overlap': 16, which means another spectrum overlaps the target spectrum.

Although it was hoped we would be 
able to separate out the different orders in an iterative way, 
that has not proved possible. 
The coincidence loss complicates matters where different orders
affect the result.

\section{How and when to use the grisms}

\subsection{Determining if a source can be observed in the UVOT grism}
\label{maglimits}
With a complicated interplay between effective area, coincidence loss in 
both background and spectrum, and incident flux, some guidance is needed 
for determining when the UVOT grisms are likely to return a useful 
spectrum. 

For faint spectra, the level of the background determines the signal to 
noise. Backgrounds can vary during the orbit, for example due to Earth shine, 
and depend also on the sky background, mainly the zodiacal light, which 
depends on the distance to the ecliptic. Some 
details were provided in \citet{breeveld10}. 
The \Sw orbit of 90 minutes limits the length of a single observation. 
Multiple exposures can be added together to improve the signal to noise 
of a weak spectrum.  For example, during the fading of SN2010aw \citep{Bayless}
the last spectrum was obtained from summing nearly 20ks of exposure.
The spectra from individual exposures can be extracted and 
summed with appropriate weights, or alternatively, the images can be aligned 
on the anchor points and summed. The success of the latter method depends on 
the accuracy of the anchor positions, and usually requires the same roll angle
so that zeroth orders of field stars align as well. For the analysis in this
paper the first method was used throughout. 

For bright sources, especially variable ones, it is 
difficult to give a limiting brightness. The limit is reached when 
the effects of coincidence can no longer be corrected.  
From Eq.~\ref{basiccoi} the determining factor for the size of 
the coincidence is the incident count rate per frame. 
An inspection of the flux for the bright WD GD153 in the \uv grism around 
2700\AA, that is at the peak of the effective area, shows 
the flux to be around $1.3\times 10^{-13}$ ergs/cm$^2$/s/\AA\  (see 
Fig.~\ref{coi-spectra-160}). 
The coincidence loss factor there is $F_{coi}=1.30$. 
In the \visg the coincidence 
loss for GD153 peaks even higher, at a factor $F_{coi}=1.70$, but there 
the continuum spectrum starts to be affected rather severely by the coincidence loss.
  
\begin{figure}
\includegraphics[width=88.0mm,angle=0]{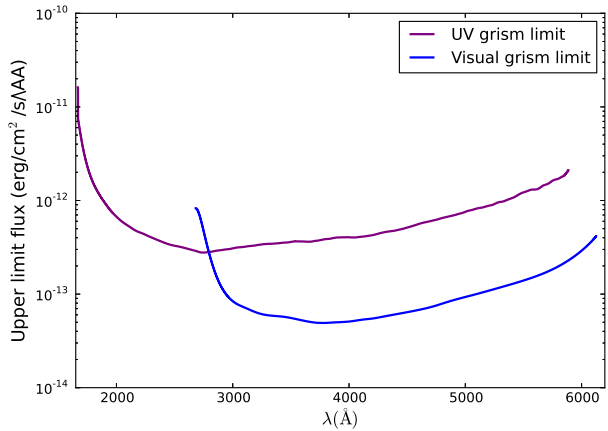}
\caption{The largest incident flux that the UVOT grisms can observe and that 
can be corrected for coincidence is shown for each detector. Larger fluxes
are not calibrated, or may damage the detector.}
\label{flux_limit_plot}
\end{figure}

Assuming an intermediate value as a reasonable limit for the 
coincidence loss of $F_{coi}^{limit}=1.5$, 
the corresponding incident count rate per bin 
per frame is 2.2. 
Using Eq.\ref{ea_eq} the following inequality follows:

\begin{equation}\label{flux_limit_eq}
F_{cal} < {{2.2 h\nu} \over {{A_{eff}} \times (\Delta{\rm \AA}/bin) } } 
erg\ cm^{-2}s^{-1}{\rm\AA}^{-1},
\end{equation}
with $h\nu$ the photon energy in $erg$.   
For the \uv and \visg we can thus graphically plot the upper limits using 
the typical effective areas and dispersion, and it is shown in Fig. \ref{flux_limit_plot}. 
The approximations made mean the 
upper limit is given to within about 30\%. If an observation is available 
for a lenticular filter, that can be used to judge whether the part of the spectrum
covered by the band is observable, see Table \ref{flux_limit_table}. 
For example, if the magnitudes in the \uv filter 
are below the limit, but the u,b,v filters are above the limit, the part 
of the spectrum above 3000\AA\  is likely to be overexposed.

\begin{table}
   \caption{Bright upper grism limit judged by photometric UVOT (Vega) magnitude.}
   \label{flux_limit_table}

   \begin{tabular}{@{}lcll}
   \hline
filter& range$^4$ &\uvG& \visG  \\
   \hline
$v$    &5010-5880&  8.78 & 11.07 \\
$b$    &3775-4930& 10.28 & 12.54 \\
$u^{3}$ &3000-3080&  9.99 & 12.00 \\
$uvw1$ &1720-3130& 10.09 &  (1)\\
$uvm2$ &1909-2650&  9.96 &  (2)\\
$uvw2$ &1610-2850&  9.32 &  (2)\\
$white$&1650-7300&  9.84 &  \\
   \hline
   \end{tabular}
   \begin{tabular}{@{}lll}
   (1) & only a small overlap to the sensitive range of \\
       & the \visg makes this band not useful. \\
   (2) & no overlap with the \visG.\\
   (3) & red leak at 4750\\
   (4) & range with effective area larger than about 2 cm$^2$\\
   \end{tabular}
\end{table}

\subsection{Using a smaller aperture}
\label{aperture_correction} 

It is sometimes advantageous to extract the spectrum with a smaller extraction width. To convert the 
measured count rates requires that the aperture correction is known for the photons missed because 
they fell outside the extraction area. 

The profile across the spectral track is mainly determined by the grism optics but in 
a small number of observations the width of the spectrum is larger than the usual width, 
perhaps because the spacecraft pointing drifted more than could be compensated 
for by the shift-and-add algorithm. 
For that reason, the aperture correction is scaled to the sigma of a fitted gaussian 
of the profile across the dispersion direction. This ensures that the energy contained within 
a certain aperture is insensitive to the small changes of the width of the spectral track. 
The measured encircled energy (see Fig.~\ref{fig_e} ) is used 
to do an aperture correction to the observed count rate. 
The coincidence loss correction is then computed as before, but after the aperture 
correction has been done.\footnote{Due to a software bug, the aperture correction 
did not work correctly for \uvotpy\  version 1.0. It has been corrected since.}

Coincidence loss also causes a \mod8 pattern, and as a result the aperture correction becomes 
inaccurate for the brighter sources. It is therefore recommended to check the results also 
by using the default aperture.

\subsection{Some tips for using the grism data}

Very rarely a data dropout causes a line of data in the image to be lost. Since this
used to cause the \uvotpy\  to crash, the data are now automatically replaced with the 
average of the next lines. 
The flagging does not keep track of that at the present time, so only inspection of the 
image can show it. A data dropout can appear to be an absorption line in a spectrum if
it is not corrected/replaced. 

The \uvot images all show a column where a few pixels are wrapped from the right hand 
side of the image to the left hand side. This can cause a few very bright pixels at the 
end of the spectrum, especially in the clocked modes. To some extent the software 
ignores this in, for example, determining the background, but in rare occasions that may
show up as a problem.

If the field as a whole has a high count rate, the \uvot data processing can not handle 
all the data coming from the CCD within one frame time.  
That means that the upper part of the image does not get stored and appears black. 
There is then a small region between the image at the bottom
and the dark top where the count rate ramps up. This is thought to be governed by the 
statistical fluctuations in the incoming rate. It seems that the bottom part 
of such an image is correctly exposed, and tests on both photometry and spectra support 
this.  

Most recent grism exposures are taken in a mode which pairs them with a 
lenticular filter. The spectra can then be extracted by 
either using the lenticular filter image to refine the astrometry or using the 
header information after a correction with {\tt uvotgraspcorr} 
which routinely is applied during ground processing. 
The larger wavelength errors when using {\tt uvotgraspcorr} are 
disappointing, since early in the mission many grism exposures were 
taken without a lenticular filter exposure. Further work may be 
able to overcome this limitation so that better wavelengths can 
be obtained for all \uvot spectra. 

Since the wavelengths are determined from the distance in pixels to 
the anchor an erroneous anchor position can lead to a distortion in the wavelength scale
of tens of angstroms at the end of the wavelength range. 
If the anchor position shift is known, this can be corrected by rederiving
the wavelengths using the dispersion relation for the spectrum given 
in the header.


\section*{acknowledgements}


Throughout the calibration many people provided feedback and helped refine 
the understanding of what was possible and what could be improved through their 
use of the grisms for observational studies. We wish to thank all of them for their 
efforts, help and patience.  A special thanks goes to all the Swift planners 
who through their effort ensured the success of this calibration.
The optical design of the grisms was by the late Richard Bingham.
We are grateful to Fred Walter, Ed Sion, and Greg Schwartz 
for sharing their HST and optical spectra, some of which were taken 
during several Swift-HST observing campaigns, which helped the calibration effort. 
This work was supported by the U.K. Space Agency
through a grant for \Sw Post Launch Support at UCL-MSSL. 
This work is sponsored at PSU by NASA contract NAS5-00136.
We acknowledge the use of data from the SIMBAD and Vizier data bases 
at the CDS in Strassbourg, 
the online WR spectra from Hamann, 
the STScI MAST and HLA archive, the ESA INES IUE archive, 
the HEASARC archives,  
and the NIST atomic data base on the WWW.
We used "Astropy", a community-developed core Python
package for Astronomy (Astropy Collaboration, 2013).

\bibliographystyle{mn}

\label{lastpage}
\end{document}